\documentclass[nofootinbib,amsmath,prd,aps,superscriptaddress,tightenlines,12pt]{revtex4}
\usepackage{graphicx}
\usepackage{epstopdf}
\usepackage{bm}
\usepackage{epsfig}
\usepackage{graphics}
\usepackage{xspace}
\usepackage{subfigure}
\usepackage{amsmath}
\usepackage{xcolor}

\usepackage{amssymb}

\def\OMIT#1{}

\newcommand{\nn}{\nonumber}

\newcommand{\bea}{\begin{eqnarray}}
\newcommand{\eea}{\end{eqnarray}}

\newcommand{\gsim}{\mathrel{\rlap{\lower4pt\hbox{\hskip1pt$\sim$}}\raise1pt\hbox{$>$}}}

\newcommand{\be}{\begin{equation}}
\newcommand{\ee}{\end{equation}}

\allowdisplaybreaks

\begin{document}

\setlength\baselineskip{17pt}


\title{\bf The 1-Jettiness DIS Event Shape at N$^3$LL+${\cal O}(\alpha_s^2)$}

\author{Haotian Cao}
\affiliation{Center of Advanced Quantum Studies, Department of Physics,
Beijing Normal University, Beijing 100875, China}

\author{Zhong-Bo Kang}
\affiliation{Department of Physics and Astronomy, University of California, Los Angeles, CA 90095, USA}
\affiliation{Mani L. Bhaumik Institute for Theoretical Physics,
University of California, Los Angeles, CA 90095, USA}
\affiliation{Center for Frontiers in Nuclear Science, Stony Brook University, Stony Brook, NY 11794, USA}
                   
\author{Xiaohui Liu}
\affiliation{Center of Advanced Quantum Studies, Department of Physics,
Beijing Normal University, Beijing 100875, China}
 \affiliation{Key Laboratory of Multi-scale Spin Physics, Ministry of Education, Beijing Normal University, Beijing 100875, China}
\affiliation{Center for High Energy Physics, Peking University, Beijing 100871, China}

\author{ Sonny Mantry}
\affiliation{Department of Physics and Astronomy, 
                   University of North Georgia,
                   Dahlonega, GA 30597, USA}



\newpage


\begin{abstract}
  \vspace*{0.3cm}

We present results for the $\tau_1$ and $\tau_{1a}$ 1-Jettiness global event shape distributions, for Deep Inelastic Scattering (DIS), at the N$^3$LL + ${\cal O}(\alpha_s^2)$
 level of accuracy. These event-shape distributions quantify and characterize the pattern of final state radiation in electron-nucleus collisions. They can be used as a probe of nuclear structure functions, nuclear medium effects in jet production, and for a precision extraction of the QCD strong coupling. The results presented here, along with the corresponding numerical codes, can be used for analyses with HERA data, in EIC simulation studies, and for eventual comparison with real EIC data.

\end{abstract}

\maketitle

\newpage

\section{Introduction}

The Electron-Ion Collider (EIC)~\cite{AbdulKhalek:2021gbh,DOE_LRP,DPAPreport2022}, to be built at the site of the Brookhaven National Laboratory, will conduct detailed studies of Quantum Chromodynamics (QCD) and the structure and dynamics of nucleons and nuclei. Some of the major goals of the EIC will be to study the origin of the nucleon mass and spin, different types of nucleon structure functions, the nuclear modification of the nucleon structure functions, the emergent properties of high density gluons at low Bjorken-$x$, and cold nuclear medium effects on the propagation color charges and jet production. To facilitate these studies, the EIC design requirements include electron-nucleon collisions at high luminosity ${\cal L}\sim 10^{33-34}$ cm$^{-2}$s$^{-1}$, a $4\pi$ hermetic detector, polarized electron and nucleus beams, collisions with a wide variety of nuclei, variable center of mass energy $\sqrt{s}\sim 20-140$ GeV, and correspondingly wide kinematic coverage in $x$ and $Q^2$, where $Q^2$ is the square of the electron momentum transfer to the nucleus. A wide range of electron-nucleus scattering observables will be studied in order to unravel these questions.

One class of observables that will be studied at the EIC are Deep Inelastic Scattering (DIS) global event shapes which characterize the pattern of final state radiation in electron-nucleus collisions. DIS event shapes were first studied~\cite{Antonelli:1999kx,Dasgupta:2001sh,Dasgupta:2001eq,Dasgupta:2002bw} more than two decades ago. The Thrust~\cite{Antonelli:1999kx} and Broadening \cite{Dasgupta:2001eq} event shapes were studied at the next-to-leading-log (NLL) level of accuracy and matched at ${\cal O}(\alpha_s)$ to fixed order results. A numerical comparison was also done \cite{Catani:1996vz,Graudenz:1997gv} against ${\cal O}(\alpha_s^2)$ results. Thrust distributions have also been measured at HERA by the H1\cite{Adloff:1997gq,Aktas:2005tz,Adloff:1999gn} and ZEUS\cite{Breitweg:1997ug,Chekanov:2002xk,Chekanov:2006hv} collaborations. Recently-proposed energy correlators~\cite{Li:2020bub,Ali:2020ksn,Li:2021txc,Liu:2022wop,Liu:2023aqb,Cao:2023rga,Devereaux:2023vjz,Andres:2023xwr,Kang:2023oqj,Cao:2023qat} further aggrandize the physics
content of the global event shapes at the EIC. 

As a generalization of the thrust observable, new event shapes were introduced and studied using the framework of 1-Jettiness~\cite{Stewart:2009yx,Stewart:2010tn}. For DIS, a dimension one 1-jettiness event shape variable $\tau_1$ was introduced in Ref.~\cite{Kang:2012zr} and resummation results were presented at the next-to-leading-logarithm (NLL) level of accuracy. These results have been extended to the NNLL~\cite{Kang:2013wca,Kang:2013nha} level of accuracy, including for electron collisions with heavier nuclei~\cite{Kang:2013wca}. These results were further improved numerically to the NNLL+${\cal O}(\alpha_s)$~\cite{Kang:2013lga} level of accuracy.  ${\cal O}(\alpha_s)$ analytic results were presented in Ref.~\cite{Chu:2022jgs} for $\tau_{1a}$ in the Breit frame and differential in $x$ and $Q^2$. In Refs.~\cite{Kang:2013nha,Chu:2022jgs} a dimensionless 1-jettiness ($\tau_{1a}$) event shape was used which is a variant of the $\tau_{1}$ definition. In Ref.~\cite{Kang:2013nha},   two other 1-jettiness event shapes were introduced and denoted as $\tau_{1b}$ and $\tau_{1c}$ and their corresponding factorization formulae and numerical results at NNLL were presented. Analytic ${\cal O}(\alpha_s)$ results for $\tau_{1b}$ were presented in Ref.~\cite{Kang:2014qba}. The $\tau_{1b}$ event shape is equivalent to the DIS thrust event shape introduced in Ref.~\cite{Antonelli:1999kx}. Numerical results for $\tau_{1b}$ have been presented~\cite{Kang:2015swk} at the N$^3$LL level of accuracy and it was recently measured using HERA data~\cite{Hessler:2021usr} and compared to ${\cal O}(\alpha_s^2)$ predictions from the program NNLOJET~\cite{Gehrmann-DeRidder:2016cdi,Currie:2016ytq,Currie:2017tpe,Gehrmann:2019hwf}. Most recently in Ref.~\cite{Knobbe:2023ehi}, the $\tau_{1b}$ groomed and ungroomed event shape distribution was studied and compared to HERA data at the NLL$'$+${\cal O}(\alpha_s)$ level of accuracy and normalized to the total ${\cal O}(\alpha_s^2)$ cross section.  Here, NLL$'$ refers to using ${\cal O}(\alpha_s)$ matrix elements, one order higher than needed for NLL resummation. Efforts toward higher logarithmic precision are under investigation recently~\cite{Bruser:2018rad, Banerjee:2018ozf, Ebert:2020unb, Baranowski:2022vcn, Baranowski:2022khd,Chen:2020dpk} and the power corrections to the class of the jettiness observables have also been studied in~\cite{Moult:2016fqy, Boughezal:2016zws,Ebert:2018lzn,Boughezal:2018mvf}. 

In this work, for the first time, we present numerical results for the $\tau_1$ and $\tau_{1a}$ event shape distributions at the N$^3$LL+${\cal O}(\alpha_s^2)$ level of accuracy. The ${\cal O}(\alpha_s^2)$ fixed order calculation, which includes up to three final state colored partons, can be implemented numerically using programs such as NLOJET++~\cite{Nagy:2005gn} where the ${\cal O}(\alpha_s)$ di-jet production in DIS is calculated, as well as the DISTRESS~\cite{Abelof:2016pby} and NNLOJET~\cite{Gehrmann-DeRidder:2016cdi,Currie:2016ytq,Currie:2017tpe,Gehrmann:2019hwf} codes where the ${\cal O}(\alpha_s^2)$ DIS single jet production is available. In this work, we make use of the NLOJET++ program to numerically implement the ${\cal O}(\alpha_s)$ and ${\cal O}(\alpha_s^2)$ fixed order contributions.
 The resummation of large Sudakov logarithms that arise in the limit of small $\tau_1$ or $\tau_{1a}$, acting effectively as a veto on additional jets beyond the leading jet, is done through a factorization theorem~\cite{Kang:2012zr,Kang:2013wca,Kang:2013nha} derived using the Soft-Collinear Effective Theory (SCET)~\cite{Bauer:2000ew,Bauer:2000yr,Bauer:2001ct,Bauer:2001yt,Bauer:2002nz,Beneke:2002ph}. The factorization formula involves a convolution product of hard, jet, soft, and beam functions that describe the physics of the hard scattering, jet production, ambient soft radiation, and initial state radiation collinear to the beam. The beam functions are further matched onto the parton distribution functions (PDFs), factoring out the dynamics of the perturbative initial state radiation from the physics of nucleon structure.  The necessary ingredients needed to carry out the Sudakov resummation at the N$^3$LL level of accuracy are now available. This includes the
fixed order ${\cal O}(\alpha_s)$~\cite{Manohar:2003vb} and ${\cal O}(\alpha_s^2)$~\cite{Idilbi_2006,Becher:2006mr} hard function, the ${\cal O}(\alpha_s)$~\cite{Bosch:2004th} and ${\cal O}(\alpha_s^2)$~\cite{Becher:2006qw,Becher:2010pd} jet function, the ${\cal O}(\alpha_s)$~\cite{Stewart:2009yx, Stewart:2010qs, Mantry:2009qz,Berger:2010xi} and ${\cal O}(\alpha_s^2)$~\cite{Gaunt:2014xga,Gaunt:2014cfa} beam functions and the ${\cal O}(\alpha_s)$~\cite{Jouttenus:2011wh} and ${\cal O}(\alpha_s^2)$~\cite{Boughezal:2015eha} soft function.  Finally, the analytic expression for the four loop cusp anomalous dimension, needed to solve the renormalization group evolution equations at N$^3$LL,  was obtained recently~\cite{Henn:2019swt,Moult:2022xzt}.


\section{Kinematics and 1-Jettiness Event Shape Observables}
We consider the electron-proton DIS process\footnote{We use slightly different notation compared to our earlier works in Refs.~\cite{Kang:2012zr,Kang:2013wca,Kang:2013lga} and introduce it here in a self-contained manner.}:
\bea
\label{process}
e^-(k) + p(P) \to e^-(k') + J + X,
\eea
where  $k^\mu, k'^\mu,$ and $P^\mu$ denote the four-momenta of the initial electron, the final electron, and the initial proton, respectively, and $J$ denotes the leading jet.  We work in the center of the mass frame. Correspondingly, definitions of $\tau_1$ and $\tau_{1a}$ given below are also in the center of mass frame. For the center of mass energy $\sqrt{s}$, the initial electron and proton momenta are given by
\bea
P^\mu &=& \frac{\sqrt{s}}{2}n^\mu, \qquad n^\mu = (1,0,0,1), \nn \\
k^\mu &=& \frac{\sqrt{s}}{2}\bar{n}^\mu, \qquad \bar{n}^\mu = (1,0,0,-1),
\eea
where we have ignored the electron and proton masses. The relevant and standard DIS kinematic variables are defined as:
\bea
s&=&(k+P)^2 , \nn \\
q&=& k -k', \nn \\
Q^2 &=& -q^2, \nn \\
x &=& \frac{Q^2}{2P\cdot q}, \nn \\
y &=& \frac{P\cdot q}{P\cdot k}, 
\eea
where $Q^2= x y s$, when we ignore the proton mass.  The dimension one DIS global event shape, $\tau_1$, is defined as
\bea
\label{tau1}
\tau_1 &=& \sum_k \text{min} \Big \{ \frac{2q_B\cdot p_k}{Q_B}, \frac{2q_J\cdot p_k }{Q_J}\Big \}, 
\eea
where the sum is over all final state particles, except the final electron.  Here $q_B^\mu$ and $q_J^\mu$ denote the beam and jet reference vectors, respectively.  The $Q_B$ and $Q_J$ are constants associated with the beam and jet sectors. The choice of these quantities is part of the definition of $\tau_1$. Thus, each final state particle with momentum $p_k$ is grouped either with the beam or jet sector according to the minimization condition in Eq.~(\ref{tau1}), and contributes accordingly to $\tau_1$. Note that the largest contributions to $\tau_1$ come from final state particles with large energies and large angles relative to both the beam and jet reference vectors. The contribution of soft particles or energetic particles closely aligned with the beam or jet axes is suppressed. In this manner, the $\tau_1$ event shape quantifies the pattern of final state radiation in electron-nucleus collisions.

For the beam sector, we work with the canonical choice
\bea
\label{eq:beam_ref_choices}
q_B = x P, \qquad Q_B=x  \sqrt{s}.
\eea
The jet reference vector $q_J^\mu$ is determined by employing a standard jet algorithm~\cite{Cacciari:2011ma} such as the anti-$k_T$, $k_T$, or Cambridge-Aachen (C/A). The jet algorithm is used to determine the leading jet and its momentum $K_J^\mu$. The transverse momentum $K_{J_T}=|\vec{K}_{J_T}|$ and rapidity $y_K$ of the leading jet is used to construct the null jet reference vector $q_J^\mu$. Accordingly, for the jet sector, we work with the canonical choice
\bea
\label{eq:jet_ref_choices}
 Q_J = 2K_{J_T}\cosh y_K , \qquad q_J= (K_{J_T}\cosh y_K, \vec{K}_{J_T},K_{J_T}\sinh y_K).
\eea

We also give results for a related event shape $\tau_{1a}$~\cite{Kang:2013nha}, corresponding to a different choice for the $Q_B$ and $Q_J$ constants in Eq.~(\ref{tau1}). It is dimensionless  and defined as
\bea
\label{tau1a}
\tau_{1a} &=& \sum_k \text{min} \Big \{ \frac{2q_B\cdot p_k}{Q^2}, \frac{2q_J\cdot p_k }{Q^2}\Big \}.
\eea

In this work, we make predictions for two types of observables. The first type of observable, studied in Refs.~\cite{Kang:2012zr,Kang:2013wca,Kang:2013lga} is differential in $(\tau_1,P_{J_T},y_J)$ 
\bea
\label{obs0}
d\sigma \left [\tau_1, P_{J_T}, y_J \right ] \equiv \frac{d^3\sigma (e^- + p \to  J + X)}{dy_J\> dP_{J_T}\>d\tau_1},
\eea
where $P_{J_T}=|\vec{P}_{J_T}|$ and $y_J$ denote the transverse momentum and rapidity, respectively of the jet $J$ and they are defined through a 1-jettiness-based algorithm. Procedurally, after using a standard jet algorithm to determine the jet reference vector $q_J$, as in Eq.~(\ref{eq:jet_ref_choices}), the 1-jettiness jet momentum is defined as~\footnote{In practice, one could directly use the leading jet momentum $K_J$ as the 1-jettiness momentum $P_J = K_J$. In the resummation region where $\tau_{1}$ is small, the difference in the definitions is power suppressed. In this work, we stick to Eq.~(\ref{eq:pjet}).}
\bea
\label{eq:pjet}
P_J &=& \sum_k p_k \>\theta (\frac{2q_B\cdot p_k}{Q_B} - \frac{2q_J\cdot p_k}{Q_J} ),
\eea
corresponding to the sum of the momenta of all final state particles grouped with the jet sector according to the minimization condition in Eq.~(\ref{tau1}). The jet transverse momentum $P_{J_T}$ and rapidity $y_J$ are constructed from this 1-jettiness jet momentum $P_J^\mu$, defined in the center of mass frame.

The second type of observable requires reconstruction of the DIS variables $(Q^2,x)$ or $(Q^2,y)$. Two examples of such an observable that we will work within this paper are given below
\bea
\label{obs1}
d\sigma \left [\tau_{1a}, Q^2, x \right ] &\equiv& \frac{d^3\sigma (e^- + p \to e^- + J + X)}{dx\> dQ^2\>d\tau_{1a}} , \nn \\
d\sigma \left [\tau_{1a}, Q^2, y \right ]  &\equiv& \frac{d^3\sigma (e^- + p \to e^- + J + X)}{dy\> dQ^2\>d\tau_{1a}}, 
\eea
where the two are related by a simple Jacobian
\bea
\label{eq:tau1a_x_vs_y}
d\sigma \left [ \tau_{1a}, Q^2, y \right ] &=& \frac{Q^2}{y^2 s}\> d\sigma \left [\tau_{1a}, Q^2, x=\frac{Q^2}{y s} \right ], \nn \\\qquad d\sigma \left [ \tau_{1a}, Q^2, x \right ] &=& \frac{Q^2}{x^2 s}\> d\sigma \left [\tau_{1a}, Q^2, y=\frac{Q^2}{x s} \right ]
\eea
where we made use of the kinematic relation $Q^2 = x y s$. 

In order to establish notation and convention, we give the explicit expression for the tree-level cross section with single boson ($\gamma^*/Z^*$)  exchange in the parton model. The relevant electromagnetic and neutral weak currents for the electron and quarks are
\bea
\label{eq:current_ops}
J_{f, \gamma}^\mu = Q_f \bar{\psi}_f \gamma^\mu \psi_f, \qquad J_{f, Z}^\mu = \bar{\psi}_f (v_f \gamma^\mu + a_f \gamma^\mu \gamma_5 )\psi_f ,
\eea
where $Q_f, v_f, a_f$ denote the electric charge, neutral weak vector charge, and neutral weak axial-vector charge, respectively, of the fermion $f$ in units of the proton charge $e$. At the tree level, ignoring hadronization effects, the final state is just a single quark or anti-quark recoiling against the final state electron. In this case, the 1-jettiness event shape vanishes so that the resulting 1-jettiness distribution will be proportional to $\delta(\tau_1)$ or $\delta(\tau_{1a})$. Of course, these distributions will be smeared once hadronization and non-perturbative soft radiation effects are included. Ignoring final state non-perturbative effects, the resulting tree level cross section for the observable differential in $(\tau_{1a}, Q^2,x)$ is
\bea
\label{eq:tree_obs1}
d\sigma_0[\tau_{1a}, Q^2,x] = \delta(\tau_{1a}) \>\sigma_0^b  \> \Big  [ \sum_q L_q f_q(x,\mu) + \sum_{\bar{q}} L_{\bar{q}} f_{\bar{q}}(x,\mu)  \Big ],
\eea
where $f_q$ and $f_{\bar{q}}$ denote the quark and anti-quark PDFs, respectively, and  $\sigma_0^b$ is given by
\bea
\sigma_0^b = \frac{2\pi \alpha_{em}^2}{Q^4} \left [ 1+ \left (1-\frac{Q^2}{x s} \right )^2  \right ] ,
\eea
and following the notation of Ref.~\cite{Kang:2013nha}, $L_q$ and $L_{\bar{q}}$ are each respectively given by
\bea
L_{q,\bar{q}} &=& Q_q^2 - \frac{2 Q_q v_q v_e}{1+m_Z^2/Q^2} + \frac{(v_q^2 + a_q^2)(v_e^2+a_e^2)}{(1+m_Z^2/Q^2)^2} \nn \\ 
&&\mp \frac{2y(2-y)}{(1-y)^2+1} \frac{a_qa_e[Q_q(1+m_Z^2/Q^2)-2v_qv_e]}{(1+m_Z^2/Q^2)^2},
\eea
where $m_Z$ denotes the $Z$-boson mass.   The tree level cross section for the observable differential in $(\tau_{1a}, Q^2,y)$ can be obtained from Eq.~(\ref{eq:tau1a_x_vs_y}) as $d\sigma_0[\tau_{1a}, Q^2,y] = Q^2/(y^2 s) \>d\sigma_0[\tau_{1a}, Q^2,x=Q^2/(y s)]$. We also give the tree level cross section for the observable differential in $(\tau_1, P_{J_T},y_J)$,  where now $P_{J_T}$ and $y_J$ become the transverse momentum and rapidity, respectively, of the final quark or anti-quark. The result is given by
\bea
\label{eq:tree_obs0}
d\sigma_0[\tau_1,P_{J_T},y_J] = \delta(\tau_1)\>\sigma_0 \> \Big  [ \sum_q L_q f_q(x_*,\mu) + \sum_{\bar{q}} L_{\bar{q}} f_{\bar{q}}(x_*,\mu)  \Big ] ,
\eea
where we have defined $\sigma_0$ and $x_*$ as
\bea
\label{eq:sig0_xstar}
\sigma_0 = 4\pi\alpha_{em}^2\frac{e^{y_J}}{\sqrt{s}\>  P_{J_T}^2} \left [1 + \left(1-\frac{P_{J_T}}{\sqrt{s}} e^{-y_J} \right )^2 \right ], \qquad x_* = \frac{\frac{P_{J_T}}{\sqrt{s}}\>e^{y_J}}{1 - \frac{P_{J_T}}{\sqrt{s}} \>e^{-y_J}}.
\eea

We note some complementary differences between the two types of 1-jettiness observables defined in Eq.~(\ref{obs0}) and Eq.~(\ref{obs1}). The observable in Eq.~(\ref{obs0}) is differential in terms of the ($P_{J_T},y_J$) variables that are typically used in the study of jets. The observables in Eq.~(\ref{obs1}) are differential in terms of the variables $(Q^2,x)$ or $(Q^2,y)$ that are typically used in the study of inclusive DIS. 

The hard scale for the DIS scattering process is set by $\mu_H \sim P_{J_{T}}$ and $ \mu_H \sim \sqrt{Q^2}$ for the observables defined in Eqs.~(\ref{obs0}) and (\ref{obs1}), respectively. In fixed order perturbation theory in $\alpha_S$,  applicable in the region $\tau_1 \sim P_{J_T}$ or $\tau_{1a} \lesssim 1$, there can be qualitatively different kinematic configurations that contribute to the two types of observables. For example, the observables in Eq.~(\ref{obs1}), where the hard scale is set by $\mu_H \sim \sqrt{Q^2}$, require the scattered electron to emerge from the primary scattering vertex with a large $Q^2$.
On the other hand, since $\mu_H\sim P_{J_T}$ for the observable in Eq.~(\ref{obs0}), it can receive contributions from the $Q^2 \to 0$ region at {\cal O}($\alpha_s^2$) and corresponds to the leading jet recoiling against hard initial state QCD radiation. 

Since the observable in Eq.~(\ref{obs0}) does not require the measurement of $(Q^2,x,y)$, it does not require a reconstruction of the momentum of the electron emerging from the primary scattering vertex. In particular, the variables $(\tau_1,P_{J_T}, y_J)$ are determined by the momenta of all the final state particles, except the final electron, that hit the detector. Thus, unlike the observables in Eq.~(\ref{obs1}), the observable in Eq.~(\ref{obs0}) is not affected by the uncertainties associated with reconstructing the true $(Q^2,x,y)$ values which can differ from the corresponding measured values due to QED radiation emitted by the electron in initial and final states~\cite{Kripfganz:1991,BLUMLEIN2003242,Afanasev:2004,Liu:2020rvc}.

Thus, the two types of 1-jettiness observables in Eqs.~(\ref{obs0}) and (\ref{obs1}) are complementary to each other and we provide results for both.



\section{1-Jettiness Spectrum}

\begin{figure}
\subfigure  {\label{fig:subfig1}\includegraphics[scale=0.2]{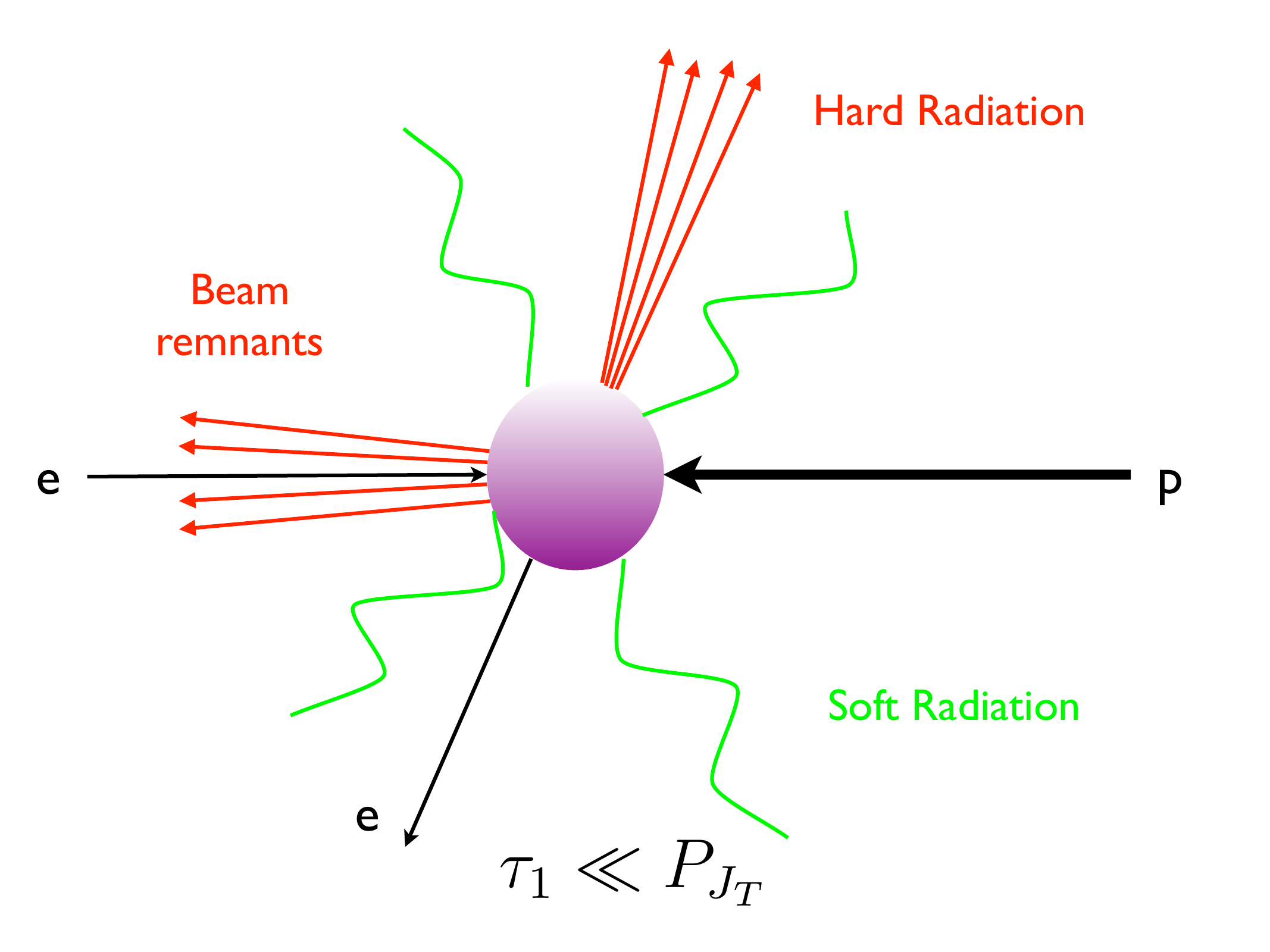}}
\subfigure {\label{fig:subfig2}\includegraphics[scale=0.2]{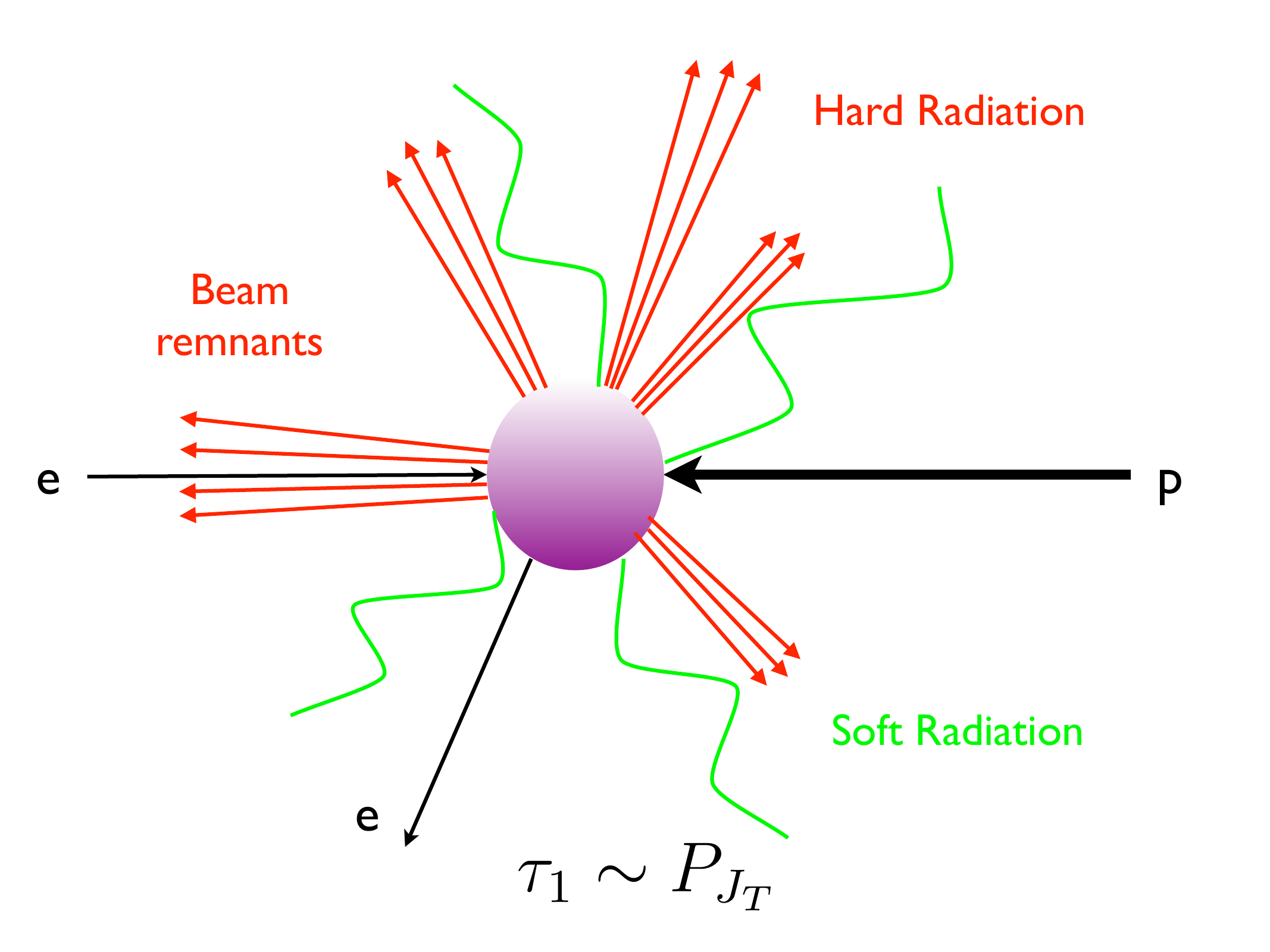}}
\caption{Schematic figure of the process $e^- + p \to e^- + J + X$ in the limit $\tau_1\ll P_{J_T}$. The restriction $\tau_1\ll P_{J_T}$ (left panel) allows only soft radiation between the beam and jet directions. In the region of large 1-jettiness $\tau_1 \sim P_{J_T}$ (right panel), additional hard radiation is allowed at wide angles from the leading jet and beam directions.}
\label{fig:process}
\end{figure}

The 1-jettiness spectrum is characterized by two distinct regions as shown in Fig.~\ref{fig:process}. The region corresponding to $\tau_1 \ll P_{J_T}$ or $\tau_{1a} \ll 1$ corresponds to the left panel of Fig.~\ref{fig:process} where the event is characterized by energetic radiation ($E\sim P_{J_T}$ or $E\sim \sqrt{Q^2}$) only along the beam or jet directions and only soft radiation ($E\sim \tau_1$ or $E\sim \tau_{1a}\sqrt{Q^2}$) at wide angles from the beam or jet directions. This can be understood from the definitions of $\tau_1$ and $\tau_{1a}$  in Eqs.~(\ref{tau1}) and (\ref{tau1a}), respectively, where it is seen that the largest contributions come from energetic final state particles at wide angles from both the beam and jet directions. On the other hand, the final state particles with momenta closely aligned with the beam or jet reference vectors $q_B^\mu$ or $q_J^\mu$,  respectively,  give small contributions.

The region of $\tau_1 \ll P_{J_T}$ or $\tau_{1a} \ll 1$  is referred to as the resummation region due to the presence of large Sudakov logarithms of the form $\alpha_s^n \ln ^{2m}(\tau_1/P_{J_T})$ or $\alpha_s^n \ln ^{2m}(\tau_{1a})$, for $m\leq n$, that arise from the small 1-jettiness restriction on final state radiation and require resummation for making accurate predictions. The small 1-jettiness restriction effectively acts as a veto on additional energetic jets at wide angles from the beam or jet references vectors, $q_B^\mu$ or $q_J^\mu$,  respectively. 

On the other hand, the region corresponding to $\tau_1 \sim P_{J_T}$ or $\tau_{1a} \lesssim 1$ corresponds to the right panel of Fig.~\ref{fig:process}, where the event is characterized by additional energetic radiation at wide angles from both the beam or leading jet directions. This corresponds to a looser veto on additional jets. In this region, the fixed order region, there are no large Sudakov logarithms so that resummation is not required and accurate predictions can be made using fixed order perturbative QCD calculations.

The resummation region, $\tau_1 \ll P_{J_T}$ or $\tau_{1a} \ll 1$, can be further classified into two sub-regions. The region $\Lambda_{QCD} \ll \tau_1 \ll P_{J_T}$ or $\Lambda_{QCD}/\sqrt{Q^2} \ll \tau_{1a} \ll 1$  corresponds to the resummation region with perturbative soft radiation. The other region, $\tau_1 \sim \Lambda_{QCD}$ or $\tau_{1a} \sim \Lambda_{QCD}/\sqrt{Q^2}$, corresponds to the resummation region with non-perturbative soft radiation. In terms of the factorization theorem, the two regions correspondingly refer to a soft function that is either perturbatively calculable or is a non-perturbative function that is typically modeled for the purposes of generating numerical results. A constraint on the non-perturbative soft function model is that it smoothly reduces to the perturbative soft function as $\tau_1$ or $\tau_{1a}$ is increased. 

The three regions of the 1-jettiness spectrum discussed above are summarized in Table~\ref{tab:1-jettiness-regions}. The complete 1-jettiness spectrum with a matching of the resummation and fixed order regions is given by the standard schematic formula
\bea
\label{eq:matched_spectrum}
d\sigma = \left [ d\sigma_{\rm resum} - d\sigma^{\rm FO}_{\rm resum}  \right ] + d\sigma^{\rm FO},
\eea
where $d\sigma_{\rm resum} $ denotes the resummed cross section in the region $\tau_1 \ll P_{J_T}$ or $\tau_{1a} \ll 1$, $d\sigma^{\rm FO}_{\rm resum}$ denotes this resummed cross section expanded to fixed order in perturbation theory, and $d\sigma^{\rm FO}$ denotes the full cross section at the same fixed order in perturbation theory. The expanded resummed cross section $d\sigma^{\rm FO}_{\rm resum}$ differs from the full fixed order cross section $d\sigma^{\rm FO}$ by terms that are non-singular in the $\tau_1\to 0$ or $\tau_{1a}\to 0$ limit. The formula in Eq.~(\ref{eq:matched_spectrum}) has the required properties for generating a smooth and continuous spectrum across the resummation and fixed order regions. In particular, we see that in the singular limit $\tau_1\to 0$ or $\tau_{1a}\to 0$, the cross section is dominated by the resummed cross section $d\sigma_{\rm resum} $  due to a cancellation between $d\sigma^{\rm FO}_{\rm resum}$ and $d\sigma^{\rm FO}$ up to suppressed non-singular terms. On the other hand, in the fixed order region $\tau_1 \sim P_{J_T}$ or $\tau_{1a} \lesssim 1$, the cross section is dominated by the full fixed order cross section $d\sigma^{\rm FO}$ due to a cancellation between $d\sigma_{\rm resum}$ and $d\sigma^{\rm FO}_{\rm resum}$ up to terms suppressed in perturbation theory. 

In the rest of the section, we discuss the features of the resummation and fixed order regions in more detail before providing numerical results.

\begin{table}[]
    \centering
    \begin{tabular}{|c|c|c|}
    \hline
      Regions & $\tau_1$ &  $\tau_{1a}$\\
       \hline
       \hline
      Resummation Region &$\tau_1 \sim \Lambda_{{\rm QCD}}$ &  $\tau_{1a} \sim \>\>\Lambda_{QCD}/\sqrt{Q^2}$ \\
       (nonperturbative soft radiation) &  &\\
       \hline
        Resummation Region   & $\>\>\Lambda_{QCD} \ll \tau_1  \ll P_{J_T}\>\>$ & $\>\>\Lambda_{QCD}/\sqrt{Q^2} \ll  \tau_{1a}  \ll 1\>\>$ \\
         (perturbative soft radiation) & & \\
         \hline
        Fixed Order Region   & $\tau_1 \sim P_{J_T}$ & $\tau_{1a} \lesssim 1$ \\
       \hline
    \end{tabular}
    \caption{The three distinct regions in the $\tau_1$ and $\tau_{1a}$ 1-jettiness spectra. }
    \label{tab:1-jettiness-regions}
\end{table}

\subsection{Resummation Region}

The resummation region, characterized by the conditions
\bea
\tau_{1} \ll P_{J_T}, \qquad \text{or} \qquad \tau_{1a} \ll 1,
\eea
allows for writing down a factorization formula that is systematically improvable, facilitates the resummation of large Sudakov logarithms, and is independent of the external jet algorithm used to determine the jet reference vector $q_J^\mu$ in Eq.~(\ref{eq:jet_ref_choices}).  For the purposes of discussing and demonstrating the jet algorithm independence, it is more convenient and natural to work with the observable $d\sigma\left [\tau_1,P_{J_T},y_J\right ]$, since it is differential in the $P_{J_T}$ and $y_J$ variables that are directly related to the properties of the leading jet.

We can understand the external jet algorithm independence in the resummation region, $\tau_1 \ll P_{J_T}$, by noting that in this region the typical event configurations look like the left panel of Fig.~\ref{fig:process}. These events are characterized by a single hard jet that is well separated from the beam region with only soft radiation between the beam and jet directions. For such events, the resulting difference between different jet algorithms just corresponds to the amount of soft radiation clustered with the jet. Only the jet mass is sensitive to the amount of soft radiation. In particular, its transverse momentum $K_{J_T}$ and rapidity $y_K$ are not affected by the soft radiation, up to power corrections in $\tau_1/P_{J_T}$. Thus, in the resummation region $\tau_1 \ll P_{J_T}$, the reference vector $q_J^\mu$ in Eq.~(\ref{eq:jet_ref_choices}) is independent of the external jet algorithm used to find the leading jet. Correspondingly, the resulting values of the 1-jettiness event shape $\tau_1$ and the 1-jettiness jet momentum $P_J^\mu$, according to Eqs.~(\ref{tau1}) and (\ref{eq:pjet}), respectively, are also independent of the external jet algorithm. Furthermore, we have $P_{J_T}=K_{J_T}$ and $y_J = y_K$, up to power corrections in $\tau_1/P_{J_T}$. Thus, in the resummation region, Eq.~(\ref{eq:jet_ref_choices}) can be written as
\bea
\label{eq:qJ_resum}
q_J \Big |_{\tau_1 \ll P_{J_T}} \simeq (P_{J_T} {\rm cosh} y_J, \vec{P}_{J_T}, P_{J_T} {\rm sinh} y_J), \qquad Q_J\simeq 2P_{J_T} {\rm cosh} y_J .
 \eea
 Thus, for a priori specified values of $P_{J_T}$ and $y_J$, we can unambiguously compute $d\sigma_{\rm resum} \left [\tau_1,P_{J_T},y_J \right ]$ using Eq.~(\ref{eq:qJ_resum}) in Eq.~(\ref{tau1}), without any reference to an external jet algorithm. 
 
Furthermore, in this resummation region where the final jet is initiated by the quark or anti-quark emerging from the hard scattering followed by a parton shower, one can associate the leading jet momentum as $P_J = q + xP$, up to power corrections from ambient soft radiation clustered with the jet. In general, there will be some uncertainty in applying this relationship arising from QED photon emissions by the initial and final electron that affects the reconstruction~\cite{Kripfganz:1991,BLUMLEIN2003242,Afanasev:2004,Liu:2020rvc}  of $q^\mu$, and correspondingly the $(Q^2,x)$ values at the primary electron scattering vertex. The identification $P_J = q + xP$ implies a simple relationship between $d\sigma_{\rm resum}\left [\tau_1,P_{J_T},y_J\right ]$ and $d\sigma_{\rm resum}\left [\tau_{1a},Q^2,y\right ]$ or $d\sigma_{\rm resum}\left [\tau_{1a},Q^2,x\right ]$ to all orders in perturbative QCD. This relationship, as derived in appendix~\ref{sec:Relationships}, is given by
\bea
\label{eq:obs_rel}
&&d\sigma_{\rm resum}\left [\tau_{1a},Q^2,y \right ] = \nn \\
&&  \frac{\sqrt{1-y}}{2 y} \>d\sigma_{\rm resum} \left [\tau_1=\sqrt{Q^2}\>\tau_{1a},P_{J_T}=\sqrt{Q^2(1-y)},y_J = \frac{1}{2}\ln\frac{Q^2(1-y)}{y^2 s} \right ] .
\eea
The factorization formula for $d\sigma_{\rm resum}\left [\tau_{1a},Q^2,x \right ]$ is then simply obtained from $d\sigma_{\rm resum}\left [\tau_{1a},Q^2,y \right ] $ using the general relation in Eq.~(\ref{eq:tau1a_x_vs_y}). One can easily check these relations for the tree level cross sections given in Eqs.~(\ref{eq:tree_obs1}) and (\ref{eq:tree_obs0}). 

Thus, in the rest of this section, when discussing the resummation region, we will primarily focus on $d\sigma_{\rm resum} \left [\tau_1,P_{J_T},y_J \right ]$ with the understanding that $d\sigma_{\rm resum}\left [\tau_{1a},Q^2,y\right ]$ and $d\sigma_{\rm resum} \left [\tau_{1a},Q^2,x \right ]$  can then be easily obtained from the relationships in Eqs.~(\ref{eq:obs_rel}) and the second line in Eq.~(\ref{eq:tau1a_x_vs_y}), respectively.

In Refs.~\cite{Kang:2012zr,Kang:2013wca}, it was shown the factorization formula for $d\sigma_{\rm resum} \left [\tau_1,P_{J_T},y_J \right ]$ has the schematic form
\bea
\label{eq:fac_schem}
d\sigma_{\rm resum} \left [\tau_1,P_{J_T},y_J \right ] &\sim &H \otimes B \otimes J \otimes {\cal S},  
\eea
where $H$ is the hard function describing the physics of the hard scattering, $B$ is the beam function~\cite{Stewart:2009yx} describing the physics of the perturbative collinear initial state radiation along the beam direction and the initial state PDF, $J$ is the quark jet function describing the physics of the collinear radiation along the jet direction, and ${\cal S}$ is the soft function describing the physics of the soft radiation throughout the event.  The beam function can be further factored  into a perturbatively calculable coefficient and initial state PDFs
\bea
B \sim {\cal I} \otimes f,
\eea
where ${\cal I}$ describes the perturbative initial state collinear radiation along the beam direction. Each of these functions in the factorization formula is sensitive to physics associated with a single energy scale so that one can minimize large logarithms by choosing the corresponding renormalization scales to have the scaling
\bea
\mu_H \sim P_{J_T}, \>\>\> \mu_J \sim \mu_B \sim \sqrt{\tau_1 P_{J_T}},  \>\>\> \mu_S \sim \tau_1.
\eea
Correspondingly, for the $\tau_{1a}$ observable, the renormalization scales chosen to minimize large logarithms have the scaling
\bea
\mu_H \sim \sqrt{Q^2}, \>\>\> \mu_J \sim \mu_B \sim \sqrt{\tau_{1a} Q^2},  \>\>\> \mu_S \sim \tau_{1a}\sqrt{Q^2}.
\eea
Using the renormalization group equations in SCET, the hard, beam, jet, and soft functions are evolved to the common scale $\mu$ at which the cross section is evaluated. In the process, large logarithms of $\tau_1/P_{J_T}$ or $\tau_{1a}$ are resummed in the corresponding resummation region $\tau_1\ll P_{J_T}$ or $\tau_{1a}\ll 1$, respectively. 


\subsection{Momentum Space Resummation Factorization Formula}

The detailed form of the factorization formula~\cite{Kang:2012zr,Kang:2013wca} in the resummation region, $\tau_1 \ll P_{J_T}$, is given by
\bea
\label{eq:factorization_resum}
d\sigma_{\rm resum} \left [\tau_1,P_{J_T},y_J \right ] &=&\sigma_0  \>H(\xi^2, \mu; \mu_H)  \int ds_J \int dt_B \> J(s_J, \mu;\mu_J){\cal S}\left(\tau_1 - \frac{t_B}{Q_B}-\frac{s_J}{Q_J}, \mu;\mu_S\right) 
\nn \\
&&\times \left [ \sum_{q} L_{q}\> B_{q}(t_B, x_*,\mu;\mu_B)  + \sum_{\bar{q}} L_{\bar{q}}\> B_{\bar{q}}(t_B, x_*,\mu;\mu_B) \right ] ,
\eea
where $\sigma_0$ and $x_*$ are given in Eq.~(\ref{eq:sig0_xstar}) and we have defined 
\bea
\label{eq:xi2}
\xi^2\equiv x_*  \sqrt{s} P_{J_T}e^{-y_J} = \frac{P_{J_T}^2}{1 - \frac{P_{J_T}}{\sqrt{s}} \>e^{-y_J}},
\eea
and field theoretic definitions of the hard ($H$), jet ($J$), beam ($B_{q,\bar{q}}$), and soft (${\cal S}$) functions can be found in Appendix A in Ref.~\cite{Kang:2013wca}.
The quark or anti-quark beam functions ($B_{q,\bar{q}}$) are matched~\cite{Stewart:2009yx} onto the PDFs as
\bea
\label{eq:beam}
B_{q,\bar{q}}(t_B, x,\mu;\mu_B) &=&\sum_{i} \int_x^1 \frac{dz}{z} {\cal I}_{({q,\bar{q}})i}\left(t_B, \frac{x}{z},  \mu;\mu_B\right) f_{i/p}(z,\mu_B),
\eea
where the ${\cal I}_{q i}$ or ${\cal I}_{\bar{q}i}$ are perturbatively calculable matching coefficients and the index $i$ runs over the possible initial parton species in the proton, including the quarks, the anti-quarks, and the gluon. The factorization formula presented in Refs.~\cite{Kang:2012zr,Kang:2013wca}, explicitly included only the case of single photon exchange in the hard scattering. The result in Eq.~(\ref{eq:factorization_resum}) is extended to also include the contribution from single $Z$-boson exchange in the hard scattering through the $L_{q,\bar{q}}$ coefficients~\cite{Kang:2013nha}. Note that the hard, jet, beam, and soft functions include their renormalization group evolution from their natural scales $\mu_H, \mu_J, \mu_B$, and $\mu_S$, respectively, to the common scale $\mu$. The PDF in Eq.~(\ref{eq:beam}) is evaluated at the $\mu_B$ scale using the standard DGLAP evolution. By charge conjugation and quark flavor symmetry of QCD, the quark jet function $J$ is the same for all light quark and anti-quark flavors so that $J_q(s_J, \mu;\mu_J) = J_{\bar{q}}(s_J, \mu;\mu_J)\equiv J(s_J, \mu;\mu_J) $ and is thus factored out of the sum over quark and anti-quark flavors. The soft function appearing in Eq.~(\ref{eq:factorization_resum}) is defined in terms of the generalized hemisphere soft function \cite{Bauer:2003di,Jouttenus:2011wh} as
\bea
\label{eq:soft-projection}
{\cal S}\left(\tau_1, \mu;\mu_S\right) &=& \int dk_B \int dk_J \>\delta (\tau_1-k_B-k_J) \>{\cal S} (k_B,k_J,\mu;\mu_S). 
\eea
The generalized hemisphere soft function ${\cal S} (k_B,k_J,\mu;\mu_S)$, appearing on the RHS above, is a function of two kinematic arguments $k_B,k_J$, corresponding to the contribution to $\tau_1$ of soft radiation grouped with the nuclear beam and jet directions respectively, as determined by the 1-jettiness algorithm used to calculate $\tau_1$ in Eq.(\ref{tau1}). 

In the region $\mu_S\sim \tau_1\sim \Lambda_{QCD}$, the soft function becomes non-perturbative and is modeled as a convolution between the perturbatively calculable partonic soft function and a phenomenological model function ($F_{\rm mod.}$) as
\bea
\label{eq:soft-model-conv}
{\cal S}(\tau_1, \mu_S) = \int du\>{\cal S}_{\rm part.}(\tau_1-u, \mu_S) \>F_{{\rm mod.}}(u),
\eea
with the normalization condition
\bea
\label{eq:Fmod_Norm}
\int du\> F_{{\rm mod.}}(u) =1.
\eea
This convolution structure ensures that the soft function reduces to the perturbative partonic soft function in the region $\tau_1 \gg \Lambda_{QCD}$, up to power corrections in $\Lambda_{QCD}/\tau_1$.  We choose a default parameterization for $F_{{\rm mod.}}(u)$ as~\cite{Kang:2012zr,Kang:2013wca,Kang:2013lga}
\bea
\label{eq:Fmod}
F_{{\rm mod.}}(u) = \frac{N(a,b,\Lambda)}{\Lambda} \left ( \frac{u}{\Lambda} \right )^{a-1} {\rm Exp} \left [-\frac{(u-b)^2}{\Lambda^2} \right ],
\eea
where $a,b,$ and $\Lambda$ are free parameters and $N(a,b,\Lambda)$ is a normalization factor that ensures the normalization constraint in Eq.~(\ref{eq:Fmod_Norm}). One might also consider analysis using shape function models that are expanded in a set of basis functions~\cite{Ligeti:2008ac,Abbate:2010xh}. In our analysis, we work with the default parameterization in Eq.~(\ref{eq:Fmod}). We note that in general, the shape function $F^{\rm mod.}(u)$ can depend on the beam and jet reference vectors used to define the 1-jettiness observable. Following the analysis in Ref.~\cite{Kang:2013nha}, in Appendix~\ref{appex:NPsoft function} we present an analytic formula in Eq.~(\ref{Fmod-beam-jet-ref-dependence-2}) that explicitly shows how the beam and jet reference vector dependence  can be incorporated into the shape function model $F^{\rm mod.}(u)$. However, since the focus of this work is on pushing the accuracy of the perturbative results, we use the simplified model in Eq.~(\ref{eq:Fmod}) which ignores the dynamical dependence on the jet reference vector. We leave a more detailed phenomenological analysis of shape function models that include this dependence for future work. We  also note that in general, the non-perturbative soft function effects will be different for the $\tau_1$ and $\tau_{1a}$ distributions. This difference can arise because of the difference in the measurement function at the operator level for the non-perturbative soft function, corresponding to the difference in the definitions of $\tau_1$ and $\tau_{1a}$, as seen in Eqs.~(\ref{tau1}) and (\ref{tau1a}), respectively. For simplicity, in this work, we choose to also implement non-perturbative effects for $\tau_{1a}$ by just using the non-perturbative model parameterization in Eq.~(\ref{eq:Fmod}), but  with appropriately different values for the $a,b,$ and $\Lambda$ parameters, and using Eq.~(\ref{eq:obs_rel}).

\subsection{Position Space Resummation Factorization Formula}

The factorization formula in  Eqs.~(\ref{eq:factorization_resum}) and (\ref{eq:beam}) can be written in terms of the Fourier transformed position space objects. The momentum space beam, jet, and soft functions are related to their position space counterparts by the Fourier transforms
\bea
\label{eq:FT_BJS}
{\cal I}_{(q,\bar{q})i}(t_B, x,\mu_B) &=& \int \frac{dy_B}{2\pi}\> e^{iy_B t_B} {\cal I}_{(q,\bar{q})i}(y_{B}, x,\mu_B),\nn \\
J(s_J,\mu_J) &=& \int \frac{dy_J}{2\pi} \> e^{iy_Js_J} J(y_J,\mu_J), \\
{\cal S}(\tau_1 , \mu_S)&=&\int \frac{dy_S}{2\pi} \> e^{iy_S\tau_1} \>{\cal S}(y_S, \mu_S). \nn
\eea
In position space, the renormalization group evolution becomes multiplicative so that the beam, jet, and soft functions can be evolved to the common scale $\mu$ from their natural scales at $\mu_B,\mu_J,$ and $\mu_S$, respectively, as
\bea
\label{eq:BJSevolPos}
{\cal I}_{(q,\bar{q})i}(y_B, x,\mu; \mu_B) &=& U_B(y_B,\mu,\mu_B) \> {\cal I}_{(q,\bar{q})i}(y_B,x,\mu_B), \nn \\
J(y_J,\mu;\mu_J) &=& U_J(y_J,\mu,\mu_J) \> J(y_J,\mu_J),  \\
{\cal S}(y_S, \mu;\mu_S) &=& U_S(y_S,\mu,\mu_S) {\cal S}(y_S, \mu_S),\nn
\eea
where $U_B(y_B,\mu,\mu_B), U_J(y_J,\mu,\mu_J),$ and $U_S(y_S,\mu,\mu_S)$ are the position space evolution factors for the beam, jet, and soft functions, respectively. Similarly, the hard function also has a multiplicative renormalization group evolution
\bea
\label{eq:hardfuncevol}
H(\xi^2, \mu; \mu_H) &=& U_H (\xi^2,\mu, \mu_H)H(\xi^2, \mu_H),
\eea
where $U_H (\xi^2,\mu, \mu_H)$ is the corresponding hard function renormalization group evolution factor. 

 Furthermore,  the momentum space convolution between the partonic soft function and the model function in Eq.~(\ref{eq:soft-model-conv}) becomes a simple product in position space
\bea
\label{eq:soft-model-conv-FT}
{\cal S}(y_\tau, \mu_S) &=& {\cal S}_{\rm part.}(y_\tau, \mu_S) \>F_{{\rm mod.}}(y_\tau),
\eea
where the position space model function is given by the Fourier transform
\bea
\label{eq:Fmod_FT}
F_{{\rm mod.}}(y_\tau) = \int du \> e^{-i y_\tau u}  \>F_{{\rm mod.}}(u). 
\eea
In terms of these position space objects, the factorization formula in Eq.~(\ref{eq:factorization_resum}) now takes the form
\bea
\label{eq:factorization_resum_FT}
d\sigma_{\rm resum} \left [\tau_1,P_{J_T},y_J \right ] &=&\sigma_0 \>U_H (\xi^2,\mu, \mu_H)H(\xi^2, \mu_H) \nn \\
&&\times \int \frac{dy_\tau}{2\pi}   e^{iy_\tau \tau_1}U_J(\frac{y_\tau}{Q_J},\mu,\mu_J) U_S(y_\tau,\mu,\mu_S) U_B(\frac{y_\tau}{Q_a},\mu,\mu_B)
\nn \\
&&\times J(\frac{y_\tau}{Q_J}, \mu_J)\>{\cal S}_{\rm part.}(y_\tau, \mu_S) \>F_{{\rm mod.}}(y_\tau) \nn \\
&&\times \Big [ \sum_{q}\sum_{i}  L_q \int_{x_*}^1 \frac{dz}{z} \> {\cal I}_{qi}\left(\frac{y_\tau}{Q_B},  \frac{x_*}{z},  \mu_B\right) f_{i/p}(z,\mu_B)  \\
&&+  \sum_{\bar{q}}\sum_{i}  L_{\bar{q}} \int_{x_*}^1 \frac{dz}{z} \> {\cal I}_{\bar{q}i}\left( \frac{y_\tau}{Q_B}, \frac{x_*}{z},  \mu_B\right) f_{i/p}(z,\mu_B) \Big ] \nn,
\eea
where all the hard, beam, jet, and soft functions are evaluated at their natural scales and the explicit renormalization group evolution factors evolve them to the common scale $\mu$. 

More details of the factorization formula in Eq.~(\ref{eq:factorization_resum_FT}) can be found in the appendices~\ref{appex:RGevol}, \ref{appex:FOfunctions}, and \ref{appex:NumFacFormula} which give explicit expressions for the various RG evolution factors up to N$^3$LL, explicit expressions for the hard, beam, jet, and soft functions up to ${\cal O}(\alpha_s^2)$, and a master factorization formula useful for numerical implementation, respectively.




\subsection{Profile functions}

As discussed in Refs.~\cite{Berger:2010xi} and \cite{Stewart:2011cf}, one must be careful in estimating the perturbative uncertainty in the matched spectrum of Eq.~(\ref{eq:matched_spectrum}). In particular, the fixed order contribution $d\sigma^{FO}$, appropriate in the fixed-order region where $\tau_1\sim P_{J_T}$ or $\tau_{1a}\sim 1$, depends on the single common scale, $\mu_{\rm FO}$. On the other hand, the resummed cross section depends on multiple scales; the hard function scale $\mu_H\sim \mu_{\rm FO}$ and the beam, jet, and soft function scales, $\mu_B,\mu_J,$ and $\mu_S$, respectively. These are the scales that correspondingly minimize large logarithms in the hard, beam, jet, and soft functions. The matched spectrum should approach the fixed order result, $d\sigma^{FO}$, in the fixed order region. This requires that resummation turns off as one approaches the fixed order region and the scales $\mu_B,\mu_J,$ and $\mu_S$ smoothly converge to $\mu_{FO} \sim \mu_{\rm H}$.  This is done by introducing profile functions~\cite{Kang:2013nha} which make the scales  $\mu_B,\mu_J,$ and $\mu_S$ functions of $\tau_1$ or $\tau_{1a}$.

We follow the parametrization of profile functions and the corresponding scale variations given in Eqs.~(201-204) of Ref.~\cite{Kang:2013nha}.  The profile functions in Ref.~\cite{Kang:2013nha} were implemented for the $\tau_{1a}$-distribution. We adapt the same parameterization for the $\tau_1$-distribution as well, but with the appropriate generalization as described below. The hard, beam, jet, and soft scales are given by
\bea
\mu_H &=& \mu_{\rm FO}\equiv \mu, \nn \\
\mu_{B,J} (x) &=& \Bigg [1 + e_{B,J} \>\theta(t_3-x) \>\left(1-\frac{x}{t_3} \right )^2 \Bigg ] \sqrt{\mu \> \mu_{\rm run}(x,\mu)}, \\
\mu_S(x) &=& \Bigg [1 + e_{S} \>\theta(t_3-x) \>\left(1-\frac{x}{t_3} \right )^2 \Bigg ] \mu_{\rm run}(x,\mu), \nn
\eea 
where the argument, $x$, of the beam, jet, and soft scale profile functions is given by
\bea
x=\tau_1/\mu \qquad {\rm or} \qquad x=\tau_{1a},
\eea
for the $\tau_1$ and $\tau_{1a}$ distributions, respectively. Similarly, the hard scale has typical size
\bea
\mu_H \sim P_{J_T}  \qquad {\rm or} \qquad \mu_H \sim \sqrt{Q^2},
\eea
for the $\tau_1$ and $\tau_{1a}$ distributions, respectively. The $e_{B,J,S}$ are parameters that can be varied to estimate the perturbative uncertainty associated with the variation of the beam, jet, and soft scales $\mu_{B,J,S}$. For $x>t_3$, all scales are set equal to the hard scale, $\mu_B=\mu_J=\mu_S=\mu$. The function $\mu_{\rm run}(x,\mu)$ is given by
\bea
\mu_{\rm run}(x,\mu) =
\begin{cases}
\mu_0 + a \>x^2/t_1,  & x \le t_1 \, \\
2  a  x + b, & t_1 \le x \le t_2 \,\\
\mu - a (x-t_3)^2/(t_3-t_2), & t_2 \le x \le t_3 \, \\
\mu, &  x > t_3 \, 
\end{cases}
\eea
where the parameters $a$ and $b$ are given by
\bea
a=\frac{\mu_0 -\mu}{t_1-t_2-t_3}, \qquad b=\frac{\mu\> t_1 -\mu_0 (t_2+t_3)}{t_1-t_2-t_3}.
\eea
We note that the profile function parameters, $a$ and $b$, above are unrelated to those that appear in the soft function model, $F_{\rm mod.} (u)$ in Eq.~(\ref{eq:Fmod}). The parameters in the profile functions are chosen~\cite{Kang:2013nha} to take on the values:
\bea
\mu_0= 2\>{\rm GeV}, \>\>\>\>\>\> t_1= \frac{3\> {\rm GeV}}{\mu},   \>\>\>\>\>\> t_2= 0.4,   \>\>\>\>\>\> t_3=0.6.
\eea
The central curves for the $\tau_1$ and $\tau_{1a}$ distributions correspond to profile functions with the choice $e_B=e_J=e_S=0$, along with $\mu = \mu_H$, where we set $\mu_H=P_{J_T}$ and $\mu_H=\sqrt{Q^2}$, respectively.  The scale variations to estimate the perturbative uncertainty are employed by varying the parameters $\mu, e_{B,J},$ and  $e_S$ in the profile functions, corresponding to varying the scales $\mu_H=\mu_{\rm FO}, \mu_{B,J},$ and $\mu_S$, respectively. The variations of the hard, beam and jet, and soft scales are respectively implemented by varying the parameters as:
\bea
\label{eq:scalevar}
{\rm Hard} \>(\mu_H):\>  \mu &=& 2^{\pm 1} Q_H,  \>\>\>\>\>\> e_{B,J} = 0, \>\>\>\>\>\>  e_S=0, \nn \\
{\rm Beam, Jet} \>(\mu_{B,J}):\>  \mu &=& Q_H,  \>\>\>\>\>\> e_{B,J} =\pm\frac{1}{3}, \pm\frac{1}{6}, \>\>\>\>\>\>  e_S=0,  \\
{\rm Soft} \>(\mu_S):\>  \mu &=& Q_H,  \>\>\>\>\>\> e_{B,J} =0, \>\>\>\>\>\>  e_S=\pm\frac{1}{3}, \pm\frac{1}{6},\nn 
\eea
where we have defined $Q_H=P_{J_T}$ or $Q_H=\sqrt{Q^2}$ for the $\tau_1$ and $\tau_{1a}$ distributions, respectively. 
\begin{figure}
    \centering
    \includegraphics[scale=0.8]{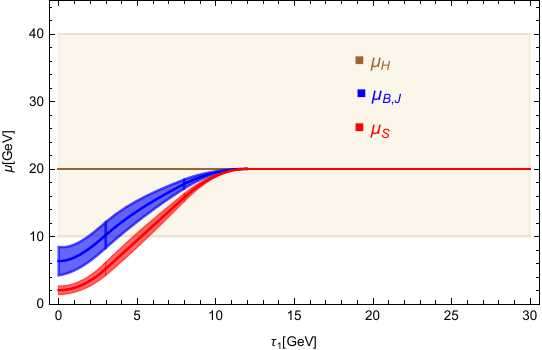}
      \includegraphics[scale=0.8]{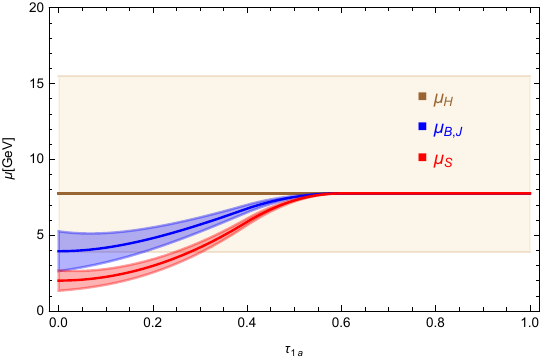}
    \caption{Profile functions for $\mu_H, \mu_{B,J},$ and $\mu_S$, along with their scale variation bands, for the $\tau_1$ (left panel) and $\tau_{1a}$ (right panel) observables. The plots correspond to the choices $Q_H=P_{J_T}=20.0$ GeV and $Q_H=\sqrt{Q^2}=\sqrt{60.0}$ GeV for the $\tau_1$ and $\tau_{1a}$ distributions, respectively. The scale variation bands are generated using the scale variations in Eq.~(\ref{eq:scalevar}).}
    \label{fig:tau1-tau1a-profiles}
\end{figure} 
Note that for the beam/jet and the soft scales there are two separate trumpet scale variations $e_{B,J,S}=\pm 1/3$ and  $e_{B,J,S}=\pm 1/6$. The scale variations for the different scales are considered one at a time and the uncertainty band is the result of adding these scale variations in quadrature.
Fig.~\ref{fig:tau1-tau1a-profiles} shows the profile functions for $\mu_H, \mu_{B,J},$ and $\mu_S$, along with their corresponding scale variations as described above, for the $\tau_1$ (left panel) and $\tau_{1a}$ (right panel). These profile function curves are for the choice $\mu_H=P_{J_T}=20.0$ GeV and $\mu_H=\sqrt{Q^2}=\sqrt{60.0}$ GeV for the $\tau_1$ and $\tau_{1a}$ distributions, respectively. We see that the profile functions smoothly connect the resummation and fixed order regions. i.e. the $\mu_H,\mu_{B,J},$ and $\mu_S$ scales have the appropriate scalings in the resummation region and smoothly converge in the fixed order region.

We note that  in the subsequent section on numerical results, for the $\tau_1$ distribution we choose $Q_H = \sqrt{\xi^2}$ in Eq.~(\ref{eq:scalevar}), instead of $Q_H=P_{J_T}$, corresponding to minimizing logarithms of the exact argument appearing in the hard function in the resummation region, as seen in Eqs.~(\ref{eq:factorization_resum}) and (\ref{eq:xi2}). This choice still has the same scaling 
$\sqrt{\xi^2}\sim P_{J_T}$ as seen in Eqs.~(\ref{eq:partMandel})
 and (\ref{eq:Q2intermsofPJTyJ}). We have also checked that both choices give consistent results.

\section{Numerical Results}

In this section, we provide numerical results up to the N$^3$LL+${\cal O}(\alpha_s^2)$ level of accuracy. For the $\tau_1$-spectrum we provide numerical results for the following choice of kinematics:
\bea
\tau_1: \qquad \sqrt{s} = 90.0\>{\rm GeV}, \qquad P_{J_T} = [20.0\>{\rm GeV},30.0\>{\rm GeV}], \qquad y_{J}=[-2.5,2.5],
\eea
corresponding to typical EIC kinematics. For the $\tau_{1a}$-spectrum we choose:
\bea
\tau_{1a}: \qquad \sqrt{s} = 319.0\>{\rm GeV}, \qquad Q^2 = [60.0\>{\rm GeV}^2,80.0\>{\rm GeV}^2], \qquad y=[0.2,0.6],
\eea
corresponding to typical HERA kinematics.  For the fixed order calculations, we use the anti-$k_T$ jet algorithm~\cite{Cacciari:2011ma} with jet radius, $R=1.0$, and numerically implement them using the NLOJET++~\cite{Nagy:2005gn} program.
\begin{figure}
    \centering
    \includegraphics[scale=0.8]{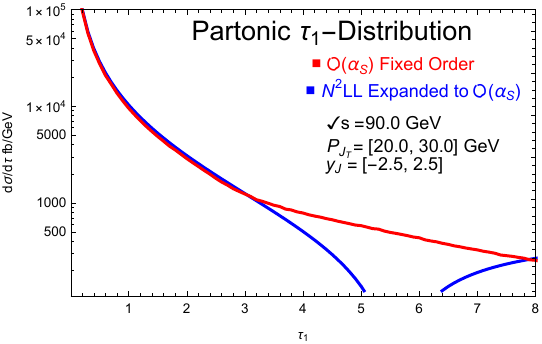}
      \includegraphics[scale=0.8]{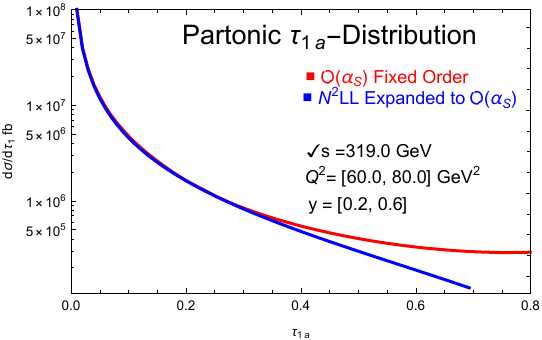}
    \caption{Comparison of the resummation expanded singular contribution (blue curve) and the full ${\cal O}(\alpha_s)$ prediction from NLOJET++ (red curve) for both $\tau_1$ (left panel) and $\tau_{1a}$ (right panel). Good agreement is observed,  validating our computational setup.  
    }
    \label{fig:tau1-tau1a-NLO-vs-N2LL}
\end{figure} 

\begin{figure}
    \centering
    \includegraphics[scale=0.8]{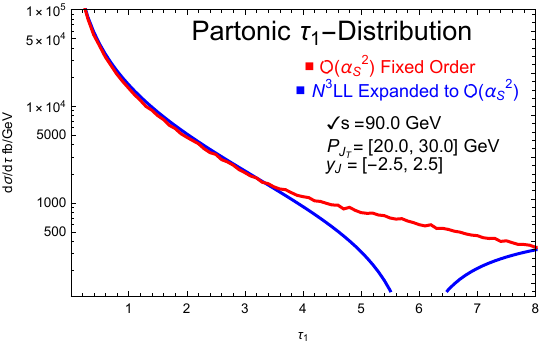}
     \includegraphics[scale=0.8]{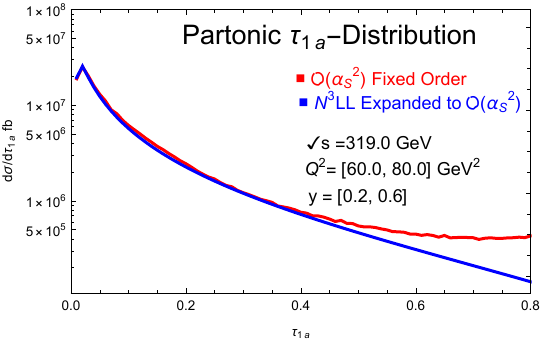}
    \caption{Comparison of the resummation expanded singular contribution (blue curve) and the full prediction from NLOJET++ up to ${\cal O}(\alpha_s^2)$ (red curve) for both $\tau_1$ (left panel) and $\tau_{1a}$ (right panel). Good agreement is observed, validating our computational setup.  
    }
    \label{fig:tau1-tau1a-NNLO-vs-N3LL}
\end{figure}

First, we provide results at the partonic level, ignoring final state hadronization effects. In Fig.~\ref{fig:tau1-tau1a-NLO-vs-N2LL}, we show the $d\sigma_{\rm resum}^{\rm FO}$ (blue) and $d\sigma^{\rm FO}$ (red) contributions to the N$^2$LL+${\cal O}(\alpha_s)$ matched cross section in Eq.~(\ref{eq:matched_spectrum}).  The left and right panels correspond to the $\tau_1$ and $\tau_{1a}$ distributions, respectively. As expected, in the small $\tau_1$ ($\tau_{1a}$) region where the fixed order result is dominated by the singular terms, $d\sigma^{\rm FO}$  at ${\cal O}(\alpha_s)$ approaches $d\sigma_{\rm resum}^{\rm FO}$ expanded to ${\cal O}(\alpha_s)$. 
In the region around $\tau_1\gtrsim 5$  GeV ($\tau_{1a}\gtrsim 0.7$), the contribution of the singular terms to the ${\cal O}(\alpha_s)$ result goes negative, and the non-singular terms in $d\sigma^{\rm FO}$ at ${\cal O}(\alpha_s)$ become important.

Similarly, in Fig.~\ref{fig:tau1-tau1a-NNLO-vs-N3LL} we show the $d\sigma_{\rm resum}^{\rm FO}$ (blue) and $d\sigma^{\rm FO}$ (red) contributions to the N$^3$LL+${\cal O}(\alpha_s^2)$ matched cross section in Eq.~(\ref{eq:matched_spectrum}).  Once again, as expected, we see that in the small $\tau_1$ ($\tau_{1a}$) region,  $d\sigma^{\rm FO}$  at ${\cal O}(\alpha_s^2)$ 
approaches $d\sigma_{\rm resum}^{\rm FO}$ expanded to ${\cal O}(\alpha_s^2)$.  Once again, in the region around $\tau_1\gtrsim 5$  GeV ( $\tau_{1a}\gtrsim 0.7$), the contribution of the non-singular terms in $d\sigma^{\rm FO}$ at ${\cal O}(\alpha_s^2)$  become important.

\begin{figure}
    \centering
    \includegraphics[scale=0.8]{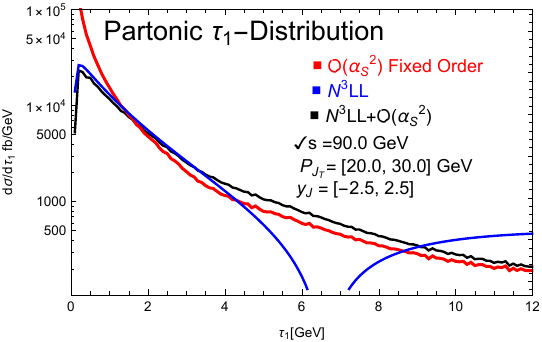}
    \includegraphics[scale=0.8]{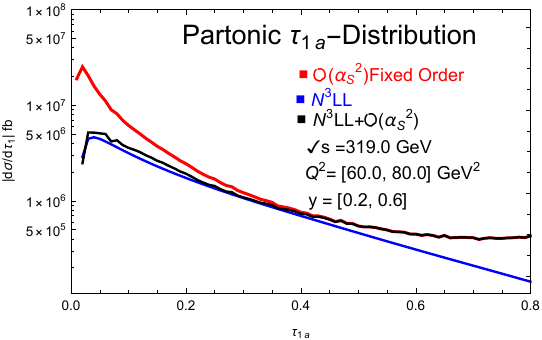}
    \caption{As seen in the left (right) panel, for the $\tau_1$ ($\tau_{1a}$) distributions of Eq.~(\ref{eq:matched_spectrum}),  the N$^3$LL+$\alpha_s^2$ (black) matched distribution approaches the N$^3$LL (blue) and ${\cal O}(\alpha_s^2)$ (red) results in the $\tau_1\ll P_{J_T}$ ($\tau_{1a}\ll 1$) and $\tau_{1a}\sim P_{J_T}$ ($\tau_{1a}\sim 1$) regions, respectively. }
    \label{fig:tau1-tau1a-N3LL-NNLO-Matched}
\end{figure}

In Fig.~\ref{fig:tau1-tau1a-N3LL-NNLO-Matched}, we show $d\sigma^{\rm FO}$ at ${\cal O}(\alpha_s^2)$ (red) , $d\sigma_{\rm resum.}$ at  N$^3$LL (blue), and the matched result, $d\sigma$, at  N$^3$LL+${\cal O}(\alpha_s^2)$ (black) for the $\tau_1$ (left panel) and $\tau_{1a}$ (right panel) distributions. We note that as expected,  the matched distribution approaches the resummation result for small $\tau_1$ or $\tau_{1a}$ and the fixed order result for large $\tau_1$ or $\tau_{1a}$.

\begin{figure}
    \centering
    \includegraphics[scale=0.8]{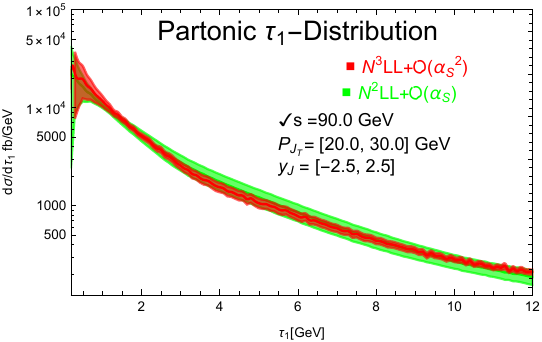}
     \includegraphics[scale=0.8]{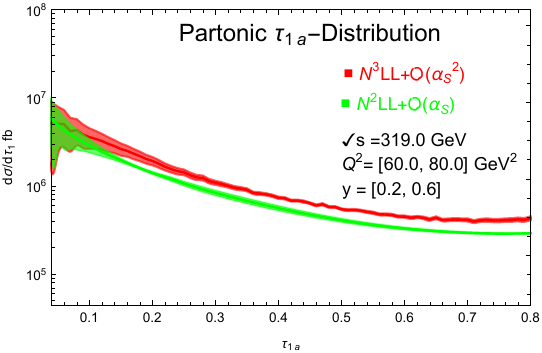}
        \caption{Partonic resummed results for $\tau_1$ (left panel) and $\tau_{1a}$ (right panel) at the N$^3$LL+${\cal O}(\alpha_s^2)$ (red) and N$^2$LL+NLO (green) levels of accuracy.}
    \label{fig:partonicscalevar}
\end{figure}

In Fig.~\ref{fig:partonicscalevar}, we show the matched $\tau_1$ (left panel) and $\tau_{1a}$ (right panel) distributions, corresponding to Eq.~(\ref{eq:matched_spectrum}), along with their scale variation bands at the N$^2$LL+${\cal O}(\alpha_s)$ (green) and N$^3$LL+${\cal O}(\alpha_s^2)$ (red) levels of accuracy. 

\begin{figure}
    \centering
    \includegraphics[scale=0.8]{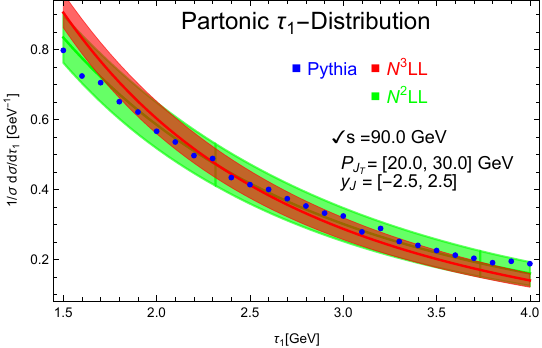}
     \includegraphics[scale=0.8]{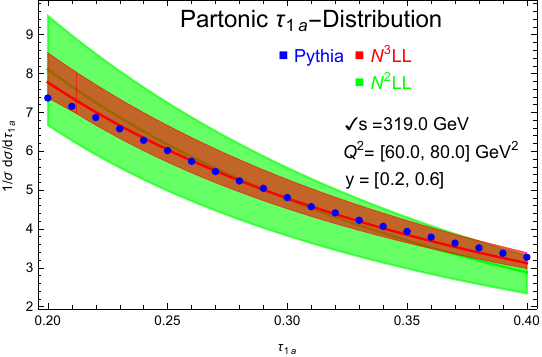}
        \caption{The $\tau_1$ (left panel) and $\tau_{1a}$ (right panel) distributions with scale variations normalized to the central curve over the displayed range, at the N$^2$LL (green) and N$^3$LL (red) level of accuracy.}
    \label{fig:Normpartonicscalevar}
\end{figure}

In Fig.~\ref{fig:Normpartonicscalevar}, we show that the $\tau_1$ (left panel) and $\tau_{1a}$ (right panel) distributions in the resummation region, normalized to the integral of the central curve over the displayed region. i.e. the curves generated through scale variation are divided by the same normalization factor used to normalize the central curve to unity over the displayed range. We see good convergence in going from the N$^2$LL to N$^3$LL resummation curves. We also display the corresponding results of Pythia8~\cite{Bierlich:2022pfr} simulations (blue dots) and find relatively good agreement.

\begin{figure}
    \centering
    \includegraphics[scale=0.8]{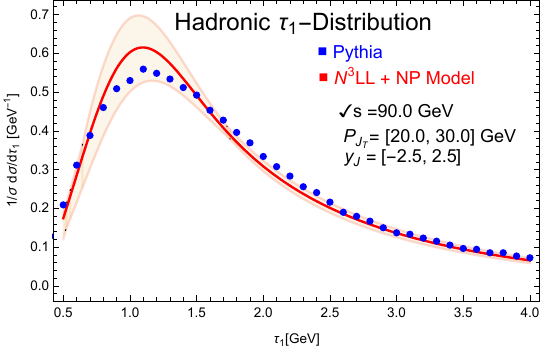}
     \includegraphics[scale=0.8]{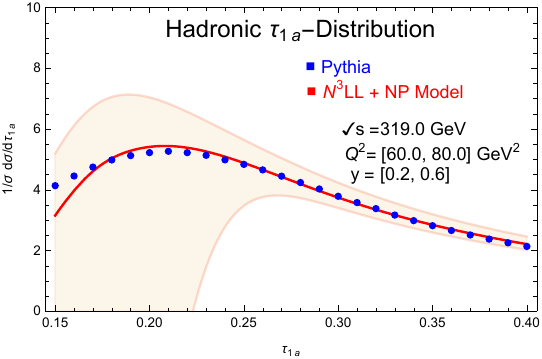}
        \caption{The $\tau_1$ (left panel) and $\tau_{1a}$ (right panel)  N$^3$LL+Soft Function Model distributions with scale variations (tan band) normalized to the central curve (red) over the displayed range, compared to Pythia data (blue dots).}
    \label{fig:HadronicSoftModelPythia}
\end{figure}

Finally, we perform a preliminary study of hadronization effects using a soft function model, following Eqs.~(\ref{eq:soft-model-conv}), (\ref{eq:Fmod_Norm}), and (\ref{eq:Fmod}). In Fig.~\ref{fig:HadronicSoftModelPythia}, we show the  $\tau_1$ (left panel) and $\tau_{1a}$ (right panel) distributions (red curve), along with their scale variation bands (tan color), in the resummation region where non-perturbative effects are important (see Table~\ref{tab:1-jettiness-regions}). The results are normalized to the integral of the central curve in the displayed region. The results are generated through a convolution of a non-perturbative soft function model with the perturbative N$^3$LL resummation curve as in Eq.~(\ref{eq:soft-model-conv}). We also show the results from Pythia8 (blue dots) with hadronization turned on. As mentioned earlier below Eq.~(\ref{eq:Fmod}), in general one expects different non-perturbative effects (soft function models) for the $\tau_1$ and $\tau_{1a}$ distributions due to the correspondingly different measurement functions at the operator level. For the $\tau_1$ distribution, the results where generated using  model parameters with values $a=1.0, b=0.45,$ and $\Lambda=0.5$ GeV in Eq.~(\ref{eq:Fmod}). For the $\tau_{1a}$ distribution, we used $a=1.0, b=0.75,$ and $\Lambda=0.5$ GeV. We see that for these choices of the soft function model parameters there is good agreement between the Pythia8 results and the theory predictions. We note that the scale variation band for the $\tau_{1a}$  distribution (right panel) in Fig.~\ref{fig:HadronicSoftModelPythia} is relatively large because of the choice of a relatively small hard scale, $\mu_H^2=Q^2=[60.0,80]$ GeV$^2$. The hard scale variation around this small central value leads to a relatively large variation in the value of the strong coupling. We have checked that at larger $Q^2$ the scale variation is much smaller and similar to what is seen for $\tau_1$, which is evaluated and varied around a larger hard scale $\mu_H \sim P_{J_T}=20.0$ GeV.

We note that these results in the non-perturbative region are only meant to demonstrate that one can easily find an appropriate soft function model to describe the Pythia8 hadronization effects. We leave a more detailed and rigorous best-fit extraction of the soft function model parameters for future work. Relatedly, there can also be important renormalon effects in the soft function~\cite{Hoang:2007vb,Abbate:2010xh}  that can affect the extraction of the soft function model parameters, and is also left for future work.

The numerical results given in this section provide a benchmark for further analyses using simulations for the proposed EIC and data collected at HERA.

\section{Conclusion}

We have provided results for the 1-Jettiness spectrum in Deep Inelastic Scattering (DIS), up to the N$^3$LL+${\cal O}(\alpha_s^2)$ level of accuracy. In particular, we considered two types of 1-Jettiness distributions, $d\sigma  \left [\tau_1, P_{J_T}, y_J \right ]$ and $d\sigma \left [ \tau_{1a}, Q^2, y \right ]$, where $\tau_1$ and $\tau_{1a}$ denote two different definitions of the 1-jettiness global event shape. We also discussed the differences and complementarity between these two types of 1-jettiness distributions. In the resummation region, corresponding to energetic final state radiation being closely aligned with either the beam or leading jet directions, a factorization framework is used and corresponding analytic formulae are provided, up to the N$^3$LL level of accuracy. In the region of very small 1-Jettiness, where the distribution becomes sensitive to the non-perturbative soft radiation throughout the event, a phenomenological model is employed to describe non-perturbative effects. In the fixed order region, corresponding to energetic final state radiation at wide angles from the beam or leading jet directions, fixed order perturbative QCD is appropriate. Fixed order results up to ${\cal O}(\alpha_s^2)$ are implemented using NLOJET++~\cite{Nagy:2005gn} program and smoothly matched with the factorization framework in the resummation region. We also provided a comparison of the theory predictions with Pythia8 simulation results, including a preliminary study of hadronization effects. These results allow for further detailed phenomenological studies of nuclear structure, nuclear medium effects, and a precision extraction of the strong coupling. The results presented here can be adapted to analyses with HERA data, ongoing EIC simulation studies, and eventual real data from the EIC.

\section*{Acknowledgement}

H.~C. and X.~L. are supported by the Natural Science Foundation of China under contract No.~12175016. Z.K. is supported by the National Science Foundation under grant No.~PHY-1945471. S.M. thanks UNG for support through the 2022 Presidential Summer Incentive Award program.

\appendix
 
 \section{Relationship Between $\tau_1$ and $\tau_{1a}$ in the Resummation Region}
 \label{sec:Relationships}
 
In this section, we derive the relationship in Eq.~(\ref{eq:obs_rel}) connecting $d\sigma_{\rm resum} \left [ \tau_{1a}, Q^2, y \right ]$ and $d\sigma_{\rm resum}  \left [\tau_1, P_{J_T}, y_J \right ]$. In the resummation region ($\tau_{1} \ll P_{J_T} \> \text{or}\>\>\>\> \tau_{1a} \ll 1$), up to power suppressed corrections, one can identify the 1-jettiness jet momentum as
 \bea
 P_J = q + x P  ,
 \eea
 corresponding to the momentum of the quark or anti-quark emerging from the hard parton-level scattering. Thus, the partonic Mandelstam variables can be expressed as
\bea
\hat{s} = (k + x P)^2, \qquad
\hat{t} = (P_J-xP)^2 = q^2 = - Q^2 ,\qquad
\hat{u} = (k-P_J)^2 ,
\eea
which can in turn be expressed in terms of $s=(k+P)^2$, $P_{J_T}$, and $y_J$ as 
\bea
\label{eq:partMandel}
\hat{s} =  x s, \qquad
\hat{t} = - Q^2 =- x \sqrt{s} P_{J_T}e^{-y_J}, \qquad
\hat{u} =  -\sqrt{s} P_{J_T} e^{y_J},
\eea
where we have ignored terms proportional to the electron or parton masses.  Furthermore, in this limit of massless electrons and partons, the partonic Mandelstam variables satisfy the constraint
 \bea
 \hat{s} + \hat{t} + \hat{u} = 0,
 \eea
 from which we can solve for the momentum fraction of the struck quark or anti-quark  to get the result
 \bea
  x &=& \frac{P_{J_T}\>e^{y_J}}{\sqrt{s} - P_{J_T} \>e^{-y_J}}. 
 \eea
 Using this result for $x$ in Eq.~(\ref{eq:partMandel}), we can write $Q^2$ and the inelasticity parameter $y=P\cdot q/P\cdot k = Q^2/(xs)$ as
 \bea
 \label{eq:Q2intermsofPJTyJ}
 Q^2 =  \frac{P_{J_T}^2}{1 - \frac{P_{J_T}}{\sqrt{s}} \>e^{-y_J}}, \qquad
 y = \frac{P_{J_T}}{\sqrt{s}} \> e^{-y_J}.
 \eea
Note that $\xi^2$ in Eq.~(\ref{eq:xi2})  is equivalent to $Q^2$  expressed in terms of $\sqrt{s},P_{J_T},$ and $y_J$, as above in Eq.~(\ref{eq:Q2intermsofPJTyJ}).  Inverting these equations, we can write $P_{J_T}$ and $y_J$ in terms of $Q^2$ and $y$ as
 \bea
 \label{eq:PJTyJtoQ2y}
 P_{J_T} = \sqrt{Q^2(1-y)}, \qquad y_J = \frac{1}{2}\ln\frac{Q^2(1-y)}{y^2 s},
 \eea
from which we obtain the Jacobian for the change of variables from $(P_{J_T},y_J)$ to $(Q^2,y)$: 
\bea
\label{eq:jacQ2yPJTyJ}
dQ^2\> dy &=& 2y \>\sqrt{\frac{Q^2}{1-y}} \>dP_{J_T}\> dy_J. 
\eea
From the definitions of $\tau_1$ and $\tau_{1a}$ in Eqs.~(\ref{tau1}) and (\ref{tau1a}), respectively, one can show that they are related to each other as
 \bea
  \label{eq:jactau1atau}
 \tau_{1a} = \frac{1}{\sqrt{Q^2}}\> \tau_{1}(Q_B\to \sqrt{Q^2}, Q_J\to \sqrt{Q^2}), \qquad  d\tau_{1a} = \frac{1}{\sqrt{Q^2}}\>d\tau_{1}.
 \eea
 Putting together Eqs.~(\ref{eq:jacQ2yPJTyJ}) and (\ref{eq:jactau1atau}), we get
 \bea
  d\tau_{1a} dQ^2\> dy &=& \frac{2y}{\sqrt{1-y}}\> d\tau_1 dP_{J_T}dy_J,
 \eea
 which along with Eq.~(\ref{eq:PJTyJtoQ2y}) gives the final result of Eq.~(\ref{eq:obs_rel}).

 

 \section{Useful Identities}
  
 The plus-distributions, ${\cal L}_n(z)$, for $n\geq 0$, are defined as
 \bea
 \label{eq:Lndist}
 {\cal L}_n(z) &\equiv& \left [ \frac{\theta(z)\ln^nz}{z}\right ]_+ =
  \lim_{\beta\to 0}\left [ \frac{\theta(z -\beta)\ln^nz}{z} +\delta(z-\beta) \frac{\ln^{n+1}\beta}{n+1}\right ],
\eea
for any dimensionless variable $z$.
Using this definition, for $\alpha \in \mathbb{R} $ and $\alpha >0$, via explicit calculation the Laplace transform of  ${\cal L}_n(z)$ is given by:
 \bea
 \int_0^\infty dz\> e^{-\alpha z}{\cal L}_n(z) =\frac{1}{n+1}\sum_{k=0}^{n+1} (-1)^{n+1-k} \binom{n+1}{k} \left (\ln\alpha\right )^{n+1-k} \int_0^\infty du\> e^{-u}\left (\ln u\right )^k .
 \eea
This  result can be analytically continued to $\alpha \to iy$, for  $y\in \mathbb{R} $, to get the Fourier transform of the $ {\cal L}_n(z)$ distributions, which are useful for computing the Fourier transforms of the beam, jet, and soft functions in Eq.~(\ref{eq:FT_BJS}). Explicit results for the cases of $n=0,1,2,$ and $3$ are:
 \bea
 \label{eq:LnFT}
 \int_0^\infty dz\> e^{-iyz} {\cal L}_0(z) &=& -L, \nn \\
\int_0^\infty dz\> e^{-iyz} {\cal L}_1(z) &=& \frac{1}{2}L^2+ \frac{\pi^2}{12}, \\
\int_0^\infty dz\> e^{-iyz} {\cal L}_2(z) &=& -\frac{1}{3}L^3 - \frac{\pi^2}{6}L-\frac{2}{3}\zeta_3, \nn \\
\int_0^\infty dz\> e^{-iyz} {\cal L}_3(z) &=& \frac{1}{4}L^4+ \frac{\pi^2}{4}L^2 +2\zeta_3L+ \frac{3\pi^4}{80}, \nn
 \eea
 where we have defined
 \bea
 L\equiv\ln(iye^{\gamma_E}),
 \eea
and Euler's constant, $\gamma_E$,  can be expressed as the definite integral:
 \bea
 \gamma_E = - \int_0^\infty du\> e^{-u}  \ln u ,
 \eea
 and $\zeta_s$ is the Riemann zeta function defined by
  \bea
 \zeta_s = \sum_{n=0}^\infty \frac{1}{n^s} = \frac{1}{\Gamma(s)}\int_0^\infty \frac{x^{s-1}}{e^x-1}dx,
 \eea
 for Re$(s) > 1$ and by analytic continuation elsewhere. Numerically, $\gamma_E\simeq 0.5772$ and $\zeta_3\simeq  1.202$. The results in Eq.~(\ref{eq:LnFT}) follow from using the identities:
 \bea
 \int_0^\infty du\> e^{-u}  \ln^2 u &=& \gamma_E^2 + \frac{\pi^2}{6}, \\
  \int_0^\infty du\> e^{-u}  \ln^3 u &=& -\gamma_E^3 -\gamma_E\frac{\pi^2}{2} -2\>\zeta_3, \nn \\
    \int_0^\infty du\> e^{-u}  \ln^4 u &=& \gamma_E^4 + \gamma_E^2 \pi^2 +8\>\gamma_E\> \zeta_3 + \frac{3\pi^4}{20} \cdot\nn 
 \eea
Another useful plus-distribution is:
\bea
\left[ \frac{\theta(x)}{x^{1+\omega}} \right ]_+&=& \lim_{\beta \to 0} \left [ \frac{\theta(x-\beta)}{x^{1+\omega}} -\delta(x-\beta) \frac{\beta^{-\omega}}{\omega}\right ],
\eea
which can be used to show:
 \bea
 \label{eq:PlusDistIdenitity}
 \big (i y e^{\gamma_E} \big )^{\omega} = \frac{e^{\omega \gamma_E}}{\Gamma(-\omega)}\int dz \>e^{-iz y}\> \left [ \frac{\theta(z)}{z^{1+\omega}} \right ]_+.
 \eea


\section{Fixed Order Results}
\label{appex:FOfunctions}
In this section, we collect results for the hard, beam, jet, and soft functions up to the ${\cal O}(\alpha_s^2)$ level of accuracy in perturbation theory. These results are needed to carry out resummation of the $\tau_{1}$- and $\tau_{1a}$-distributions at the N$^3$LL level of accuracy, using Eqs.~(\ref{eq:factorization_resum_FT}) and (\ref{eq:obs_rel}), respectively.

\subsection{Hard Function}
The hard function is given by
\bea
H(Q^2,\mu) &=& |C(Q^2,\mu)|^2 ,
\eea
where $C(Q^2,\mu)$ is the Wilson coefficient that arises from matching the QCD current operators in Eq.~(\ref{eq:current_ops}) onto the corresponding SCET current operators. The fixed order perturbvative expansion of the $C(Q^2,\mu)$ Wilson coefficient is expressed as:
\bea
C(Q^2,\mu) =\sum_{n=0}^\infty \left [\frac{\alpha_s(\mu)}{4\pi} \right ]^n  C^{(n)}
\eea
The result for the Wilson coefficient is known up to ${\cal O}(\alpha_s^2)$\cite{Idilbi_2006,Becher:2006mr}:
\bea
C^{(0)} &=& 1, \nn \\
C^{(1)} &=& C_F \big (- L^2 + 3 L -8 +\frac{\pi^2}{6}\big ), \,  \\
C^{(2)} &=& C_F \big (C_F H_F + C_A H_A + T_F n_f H_f\big ), \nn 
\eea
where $L= \ln \frac{Q^2}{\mu^2}$ and the $H_{F,A,f}$ coefficients are defined as
\bea
H_F &=& \frac{L^4}{2} - 3L^3 +\left (\frac{25}{2}-\frac{\pi^2}{6} \right )L^2 +\left (-\frac{45}{2}-\frac{3\pi^2}{2}+24\zeta_3 \right ) L + \frac{255}{8} + \frac{7\pi^2}{2}-\frac{83\pi^4}{360}-30\zeta_3, \nn \\
H_A &=& \frac{11}{9}L^3 +\left ( -\frac{233}{18}+\frac{\pi^2}{3}\right ) L^2+\left ( \frac{2545}{54}+\frac{11\pi^2}{9}-26\zeta_3\right ) L -\frac{51157}{648} -\frac{337\pi^2}{108} + \frac{11\pi^4}{45} +\frac{313}{9}\zeta_3, \nn \\
H_f&=& -\frac{4}{9}L^3 +\frac{38}{9}L^2 + \left (-\frac{418}{27} -\frac{4\pi^2}{9}\right ) L +\frac{4085}{162} + \frac{23\pi^2}{27} + \frac{4}{9}\zeta_3
\eea
The hard function can be can be correspondingly expressed as
\bea
\label{eq:Hexp}
H(Q^2,\mu) &=& |C(Q^2,\mu)|^2 = \sum_{n=0}^\infty \left [\frac{\alpha_s (\mu)}{4\pi} \right ]^n  H^{(n)},
\eea
where the coefficients $H^{(n)}$, expressed in terms of the $C^{(n)}$ coefficients, up to ${\cal O}(\alpha_s^2)$ are given by:
\bea
\label{H(n)}
H^{(0)} &=& 1, \nn \\
H^{(1)} &=& 2C^{(1)}\nn \\
H^{(2)} &=& 2C^{(2)} + (C^{(1)})^2 
\eea

\subsection{Soft Function}
The  fixed order perturbative expansion of the soft function in momentum space can be expressed as:
\bea
\label{eq:Sexpmomspace}
S(\tau_1,\mu_S) &=& \sum_{n=0}^\infty \left [\frac{\alpha_s (\mu_S)}{4\pi} \right ]^n  S^{(n)}(\tau_1, \mu_S).
\eea
The results  up to ${\cal O}(\alpha_s^2)$~\cite{Jouttenus:2011wh,Boughezal:2015eha}  are given by:
\bea
\label{eq:SmomspaceNNLO}
S^{(0)} &=& \delta(\tau_1), \nn  \\
S^{(1)} &=& C_F\left [ \frac{\pi^2}{3}  \delta(\tau_1) - \frac{16}{\tilde{\mu}}{\cal L}_1(\tau_1/\tilde{\mu}) \right ],  \nn \\  
S^{(2)} &=& 4 C_F^2 \Bigg [
         				-\frac{3\pi^4}{40} \delta(\tau_1) 
                                         + 64 \zeta_3 \frac{1}{\tilde{\mu}}{\cal L}_0(\tau_1/\tilde{\mu})
                                         - 12 \pi^2 \frac{1}{\tilde{\mu}}{\cal L}_1(\tau_1/\tilde{\mu})
                                         + 32 \frac{1}{\tilde{\mu}}{\cal L}_3(\tau_1/\tilde{\mu})
                                    \Bigg ] \nn \\ 
            &+& 4 n_f C_F \left [ 
            				\frac{2\zeta_3}{9} \delta(\tau_1)
             				-\frac{\pi^2}{9} \frac{1}{\tilde{\mu}}{\cal L}_0(\tau_1/\tilde{\mu})
				        + \frac{8}{3} \frac{1}{\tilde{\mu}}{\cal L}_2(\tau_1/\tilde{\mu})
                                       \right ]  \\
            &+& 4 C_A C_F  \Bigg [     
                                             \left (\frac{\pi^4}{240}-\frac{11\zeta_3}{9} 
                                                  -\frac{535}{81}-\frac{335\pi^2}{216}+\frac{17\pi^4}{144}+\frac{341\zeta_3}{18}
                                             \right )\delta(\tau_1) \nn \\      
                    			  &&+ \left ( \frac{11\pi^2}{18}+\frac{16\zeta_3}{3} +\frac{404}{27} 
			  				-\frac{11 \pi^2}{6} -\frac{58\zeta_3}{8} \right ) 
						\frac{1}{\tilde{\mu}} {\cal L}_0(\tau_1/\tilde{\mu}) \nn \\
			                   &&+\left (\frac{16\pi^2}{3}-\frac{268}{9} -4\pi^2\right )
			                    \frac{1}{\tilde{\mu}}{\cal L}_1(\tau_1/\tilde{\mu})
			                    +\left ( \frac{88}{3}-\frac{44}{3}\right ) 
			                       \frac{1}{\tilde{\mu}}{\cal L}_2(\tau_1/\tilde{\mu})
 				       \Bigg ]\nn  \\     
		 &+& 4 C_F n_f T_R  \Bigg [ 
		               		\left (\frac{20}{81} + \frac{37\pi^2}{54}- 
				                 \frac{62\zeta_3}{9} \right )\delta(\tau_1) 
				      + \left ( \frac{2\pi^2}{3}-\frac{112}{27}\right )
				       \frac{1}{\tilde{\mu}} {\cal L}_0(\tau_1/\tilde{\mu}) \nn \\
				   &+& \frac{80}{9} \frac{1}{\tilde{\mu}} {\cal L}_1(\tau_1/\tilde{\mu})  
				         -  \frac{32}{3} \frac{1}{\tilde{\mu}} {\cal L}_2(\tau_1/\tilde{\mu}) 
		                        \Bigg ],	\nn	                                          
\eea
where we have defined the scale
\bea
\label{eq:tildemuS}
\tilde{\mu} &=& \mu_S \>r_S, \qquad r_S =\sqrt{\frac{2 q_B\cdot q_J}{Q_B Q_J}}.
\eea
The fixed order expansion of the soft function in position space can be obtained from the momentum space result in Eq.~(\ref{eq:Sexpmomspace}) by inverting the Fourier transform relation in Eq.~(\ref{eq:FT_BJS}). The corresponding perturbative expansion in position space can be expressed as:
\bea
\label{eq:Sexp}
{\cal S}(y_\tau,\mu_S) &=& \sum_{n=0}^\infty \left [\frac{\alpha_s (\mu_S)}{4\pi} \right ]^n  {\cal S}^{(n)}(y_\tau,\mu_S),
\eea
where the ${\cal S}^{(n)}(y_\tau,\mu_S)$ coefficients are functions of $\tilde{L}_S=\ln(iy_\tau\tilde{\mu}e^{\gamma_E})$. This is apparent through the identities in Eq.~(\ref{eq:LnFT}) for the Fourier transforms of the $1/\tilde{\mu}\>\>\>{\cal L}_{n}(\tau_1/\tilde{\mu})$ distributions that appear in  Eq.~(\ref{eq:SmomspaceNNLO}). Using Eq.~(\ref{eq:tildemuS}), we can write the useful relation
\bea
\tilde{L}_S = L_S + \ln r_S,
\eea
where we  have defined
\bea
\label{eq:LS}
L_S= \ln (i y_{\tau} \mu_S e^{\gamma_E}).
\eea

\subsection{Jet Function}

The fixed order expansion of jet function in position space is parameterized as  
\bea
\label{eq:Jexp}
J(y_J,\mu_J) &=& \sum_{n=0}^\infty \left [ \frac{\alpha_s (\mu_J)}{4\pi} \right ]^n  J^{(n)}. 
\eea
The results are known up to ${\cal O}(\alpha_s^2)$~\cite{Becher:2006qw, Idilbi_2006,Becher:2006mr}: 
\bea
\label{eq:Jcoeff}
J^{(0)} &=& 1, \nn \\
J^{(1)} &=& C_F \left ( 2 L_J^2+3 L_J +7 -\frac{2\pi^2}{3}\right ),  \\
J^{(2)} &=& C_F\left (C_F J_F + C_A J_A + T_F n_f J_f\right), \nn
\eea
and $J_F,J_A,$ and $J_f $ coefficients are defined as:
\bea
\label{eq:Jcoeff2}
J_F & =& 2L_J^4 -6L_J^3+\left ( \frac{37}{2} -\frac{4\pi^2}{3}\right )L_J^2 + \left ( -\frac{45}{2} + 4\pi^2 -24 \zeta_3\right ) L_J+  \frac{205}{8} -\frac{97\pi^2}{12} + \frac{61\pi^4}{90} - 6 \zeta_3 , \nn \\
J_A &=& -\frac{22}{9} L_J^3+ \left (\frac{367}{18} - \frac{2\pi^2}{3} \right ) L_J^2 + \left (-\frac{3155}{54} + \frac{11\pi^2}{9} + 40\zeta_3 \right ) L_J \nn \\
&+& \frac{53129}{648} - \frac{155\pi^2}{36} - \frac{37\pi^4}{180} -18\zeta_3 , \nn \\
J_f &=& \frac{8}{9} L_J^3 - \frac{58}{9} L_J^2 + \left ( \frac{494}{27} -\frac{4\pi^2}{9} \right )L_J -\frac{4057}{162} + \frac{13\pi^2}{9},
\eea
where the logarithm, $L_J$, is defined as
\bea
\label{eq:LJ}
L_J= \ln (i y_J \mu_J^2 e^{\gamma_E}).
\eea

\subsection{Beam Function}

The beam function is given by the convolution 
\bea
B_i(t,x,\mu_B) = \sum_j \int_x^1 \frac{dz}{z} {\cal I}_{ij}(t,z,\mu_B) f_j(\frac{x}{z},\mu_B),
\eea
where the matching coefficients have the perturbative expansion
\bea
\label{eq:Iijexpt}
{\cal I}_{ij}(t,x,\mu_B) = \sum_n^\infty \left [ \frac{\alpha_s(\mu_B)}{4\pi}\right ]^n {\cal I}_{ij}^{(n)}(t,x,\mu_B)
\eea
Up to ${\cal O}(\alpha_s^2)$~\cite{Stewart:2010qs,Gaunt:2014xga}, the expressions for the matching coefficients are given by
\bea
\label{eq:IijFO}
{\cal I}_{ij}^{(0)}(t,x,\mu_B) &=& \delta(t) \delta_{ij} \delta(1-x), \nn \\
{\cal I}_{ij}^{(1)}(t,x,\mu_B) &=& \frac{1}{\mu_B^2} {\cal L}_1(t/\mu_B^2) \Gamma_0^i \delta_{ij} \delta(1-x) + \frac{1}{\mu_B^2} {\cal L}_0(t/\mu_B^2) \left [ -\frac{\gamma_{B0}^i}{2} \delta_{ij} \delta(1-x) + 2 P_{ij}^{(0)}(x)\right ]\nn \\
&+& \delta(t) 2I_{ij}^{(1)}(x), \nn \\
{\cal I}_{ij}^{(2)}(t,x,\mu_B) &=& \frac{1}{\mu_B^2} {\cal L}_3(t/\mu_B^2)\frac{(\Gamma_0^i)^2}{2}\delta_{ij}\delta(1-x) \nn \\
&+& \frac{1}{\mu_B^2} {\cal L}_2(t/\mu_B^2) \Gamma_0^i  \Bigg \{-\left(\frac{3}{4}\gamma_{B0}^i + \frac{\beta_0}{2} \right) \delta_{ij}\delta(1-x) + 3 P_{ij}^{(0)}(x)\Bigg \}\nn \\
&+& \frac{1}{\mu_B^2} {\cal L}_1(t/\mu_B^2) \Bigg \{ \left ( \Gamma_1^i - \frac{\pi^2}{6} +\left( \Gamma_0^i\right )^2\frac{\left( \gamma_{B0}^i\right )^2}{4} + \frac{ \gamma_{B0}^i\beta_0}{2}\right ) \delta_{ij}\delta(1-x) + 2\Gamma_0^i I_{ij}^{(1)}(x) \nn \\
&-&2\left (  \gamma_{B0}^i+\beta_0\right) P_{ij}^{(0)}(x) + 4\sum_k \int_x^1 \frac{dz}{z} P_{ik}^{(0)}(z)P_{kj}^{(0)}(\frac{x}{z})\Bigg \}    \nn \\
&+&  \frac{1}{\mu_B^2} {\cal L}_0(t/\mu_B^2)\Bigg \{\left (\left (\Gamma_0^i  \right )^2 \zeta_3 + \Gamma_0^i \gamma_{B0}^i \frac{\pi^2}{12} - \frac{\gamma_{B1}^i}{2} \right )\delta_{ij}\delta(1-x) -  \Gamma_0^i \frac{\pi^2}{3} P_{ij}^{(0)}(x) \nn \\
&-&\left (\gamma_{B0}^i + 2\beta_0 \right ) I_{ij}^{(1)}(x) + 4\sum_k \int_x^1 \frac{dz}{z}I_{ik}^{(1)}(z)P_{kj}^{(0)}(\frac{x}{z}) + 4 P_{ij}^{(1)}(x)\Bigg \} \nn \\
&+& \delta(t) 4I_{ij}^{(2)}(x)
\eea
The quark flavor diagonal and universal structure of QCD interactions results in two distinct types of non-zero matching coefficients for each quark flavor $q$, denoted by ${\cal I}_{qq}(t,x,\mu_B)$ and ${\cal I}_{qg}(t,x,\mu_B)$.  Explicit expressions for the functions $I_{ij}^{(1)}(x)$ and $I_{ij}^{(2)}(x)$ are quite long and can be found in Ref.~\cite{Gaunt:2014xga}. Similarly, one can also find explicit expressions for the splitting functions $P_{ij}^{(0)}(x)$ and $P_{ij}^{(1)}(x)$ in Ref.~\cite{Gaunt:2014xga}.
The position space matching coefficients can be obtained by inverting the Fourier transform relation in Eq.~(\ref{eq:FT_BJS}), and making use of the identities Eq.~(\ref{eq:LnFT}) for the Fourier transforms of the $1/\mu_B\>{\cal L}_{n}(t/\mu_B^2)$ distributions that appear in  Eq.~(\ref{eq:IijFO}). The corresponding perturbative expansion in position space can be expressed as:
\bea
\label{eq:Iijexpy}
{\cal I}_{ij}(y_B,x,\mu_B) = \sum_{n=0}^\infty \left [\frac{\alpha_s(\mu_B)}{4\pi} \right ]^n  {\cal I}_{ij}^{(n)}(y_B,x,\mu_B),
\eea
where the $ {\cal I}_{ij}^{(n)}(y_B,x,\mu_B)$ coefficients will be functions of
\bea
\label{eq:LB}
L_B &=& \ln(iy_B\mu_B^2e^{\gamma_E}). 
\eea


\section{Renormalization group evolution}
\label{appex:RGevol}
In this section, we collect useful results needed to carry out resummation of the $\tau_{1}$- and $\tau_{1a}$-distributions at the N$^3$LL level of accuracy, using Eqs.~(\ref{eq:factorization_resum_FT}) and (\ref{eq:obs_rel}), respectively. In particular, we collect the results for the RG evolution factors appearing in Eq.~(\ref{eq:factorization_resum_FT}).

\subsection{Hard function}
The RG evolution equation for the hard function is given by
\bea
\label{eq:HRG}
\mu \frac{d}{d\mu} H(Q^2,\mu) &=& \gamma_H \>H(Q^2,\mu),
\eea
where the anomalous dimension $\gamma_H$ is given by
\bea
\label{eq:Hanom}
\gamma_H &=& \gamma_c + \gamma_c^*,
\eea
where $\gamma_c$ is the anomalous dimension of the Wilson coefficient $C(Q^2,\mu)$ which satisfies $H(Q^2,\mu) =  | C(Q^2,\mu) |^2$. 
The general form~\cite{Becher:2009cu,Becher:2009qa} of the anomalous dimension is given by
\bea
\label{eq:Canom}
\gamma_c &=& \sum_{(i,j)} \frac{T_i\cdot T_j}{2}\gamma_{\text{cusp}}(\alpha_s) \ln \frac{\mu^2}{-s_{ij}} + \sum_i \gamma^i(\alpha_s),
\eea
where $s_{ij}=2\sigma_{ij}p_i\cdot p_j+ i0$ and $\sigma_{ij}=+1$ if the momenta $p_i$ and $p_j$ are both incoming or outgoing and $\sigma_{ij}=-1$ otherwise. The sum over $i,j$ run over the external partons of the corresponding SCET operator and $(i,j)$ denotes unordered tuples of distinct parton indices.  In our case, the SCET operator is just the photon or Z-boson current operator involving two quarks or antiquarks as the external partons.

 $\gamma_{\text{cusp}}$ is related to the cusp anomalous dimension in the fundamental and adjoint representations $\Gamma_{\text{cusp}}^F(\alpha_s)$ and $\Gamma_{\text{cusp}}^A(\alpha_s)$ respectively as
\bea
\gamma_{\text{cusp}}(\alpha_s)=\frac{\Gamma_{\text{cusp}}^F(\alpha_s)}{C_F} &=& \frac{\Gamma_{\text{cusp}}^A(\alpha_s)}{C_A}.
\eea
For example, $\Gamma_{\text{cusp}}^F(\alpha_s)$ and $\Gamma_{\text{cusp}}^A(\alpha_s)$ correspond to the case with all external lines being quarks or antiquarks and all external lines being gluons, respectively. The cusp and non-cusp anomalous dimensions and the beta function have expansions in $\alpha_s$ given by
\bea
\label{eq:pertexpanomdim}
\gamma_{\text{cusp}} [\alpha_s] &=& \sum_{n=0}^\infty \Big (\frac{\alpha_s}{4\pi}\Big )^{n+1}\gamma_n^{\text{cusp}},\nn \\
 \gamma^{q,g}[\alpha_s] &=& \sum_{n=0}^\infty \Big (\frac{\alpha_s}{4\pi}\Big )^{n+1}\gamma_n^{q,g},  \\
 \beta[\alpha_s] &=& -2\alpha_s \sum_{n=0}^\infty \Big (\frac{\alpha_s}{4\pi}\Big )^{n+1}\beta_n. \nn
\eea

For N$^3$LL resummation of $\tau_1$ and $\tau_{1a}$ we need $\gamma_{\text{cusp}}$, $\gamma^{q}$, and $\beta$ to 4-loops, 3-loops, and 4-loops respectively along with ${\cal O}(\alpha_s^2)$ PDFs. Here $\gamma^{q}$ denotes the non-cusp anomalous dimension contribution form light quarks or antiquarks. 

The coefficients of $\gamma_{\rm cusp}$ in Eq.~(\ref{eq:pertexpanomdim}), up to four loops~\cite{Korchemsky:1987wg,Moch:2004pa,Henn:2019swt},  are given by
\bea
\gamma_0^{\text{cusp}} &=& 4 , \nn \\
\gamma_1^{\text{cusp}}&=& 4  \Big [ \Big ( \frac{67}{9}-\frac{\pi^2}{3}\Big ) C_A-\frac{20}{9}T_Fn_f\Big ], \nn \\
\gamma_2^{\text{cusp}} &=& 4  \Big [ C_A^2 \Big (\frac{245}{6} -\frac{134\pi^2}{27}+ \frac{11\pi^4}{45} + \frac{22}{3}\zeta_3 \Big )+C_A T_F n_f\Big (-\frac{418}{27}+ \frac{40\pi^2}{27}-\frac{56}{3}\zeta_3 \Big )\nn \\
&+& C_FT_Fn_f\Big ( -\frac{55}{3} + 16 \zeta_3\Big ) - \frac{16}{27}T_F^2n_f^2\Big ], \\
\gamma_3^{\text{cusp}} &=& 256  \Big [ C_A^3 \Big ( \frac{1309\zeta_3}{432}
-\frac{11\pi^2\zeta_3}{144}-\frac{\zeta_3^2}{16}-\frac{451\zeta_5}{288} + \frac{42139}{10368}-\frac{5525\pi^2}{7776}+\frac{451\pi^4}{5760}-\frac{313\pi^6}{90720} \Big ) \nn \\
&+& n_fT_FC_A^2 \Big (-\frac{361\zeta_3}{54} +\frac{7\pi^2\zeta_3}{36}+\frac{131\zeta_5}{72} 
-\frac{24137}{10386}+\frac{635\pi^2}{1944}-\frac{11\pi^4}{2160}\Big )\nn \\
&+& n_f T_F C_F C_A \Big ( \frac{29\zeta_3}{9}-\frac{\pi^2\zeta_3}{6}+\frac{5\zeta_5}{4}-\frac{17033}{5184}+\frac{55\pi^2}{288}-\frac{11\pi^4}{720}\Big ) \nn \\
&+& n_f T_F C_F^2 \Big (\frac{37\zeta_3}{24}-\frac{5\zeta_5}{2}+\frac{143}{288} \Big ) + n_f^2T_F^2C_A\Big ( \frac{35\zeta_3}{27}-\frac{7\pi^4}{1080} - \frac{19\pi^2}{972} + \frac{923}{5184}\Big ) \nn \\
&+&n_f^2T_F^2C_F\Big ( -\frac{10\zeta_3}{9}+\frac{\pi^4}{180}+\frac{299}{648}\Big ) + n_f^3T_F^3 \Big ( -\frac{1}{81}+ \frac{2\zeta_3}{27}\Big ) \nn \\
&+& \frac{d_F^{abcd}d_A^{abcd}}{C_FN_c} \Big ( \frac{\zeta_3}{6} -\frac{3\zeta_3^2}{2}+
\frac{55\zeta_5}{12}-\frac{\pi^2}{12}-\frac{31\pi^6}{7560}\Big )
+n_f\frac{d_F^{abcd}d_F^{abcd}}{C_F N_c}\Big ( \frac{\pi^2}{6}-\frac{\zeta_3}{3}-\frac{5\zeta_5}{3}\Big )
\Big ] .\nn
\eea
The coefficients of $\gamma^q$ in Eq.~(\ref{eq:pertexpanomdim}), up to three loops~\cite{Moch:2005id,Becher:2006mr},  are given by
\bea
\label{eq:gamqpertexpN3LL}
\gamma_0^q &=& -6C_F , \nn\\
\gamma_1^q &=& C_F^2 (-3+4\pi^2-48\zeta_3) +  C_FC_A (-\frac{961}{27}-\frac{11\pi^2}{3}+52 \zeta_3) +  C_FT_F n_f (\frac{260}{27}+\frac{4\pi^2}{3}),\nn \\
\gamma_2^q &=& C_F^3(-29-6\pi^2-\frac{16\pi^4}{5}-136\zeta_3
                                +
                                \frac{32\pi^2\zeta_3}{3}+480\zeta_5)\nn \\
                     &+&C_F^2C_A(-\frac{151}{2}+\frac{410\pi^2}{9}+\frac{494\pi^4}{135}-\frac{1688\zeta_3}{3.0}
                                -\frac{16\pi^2\zeta_3}{3}-240\zeta_5)\nn \\
                     &+&C_FC_A^2(-\frac{139345}{1458}-\frac{7163\pi^2}{243}-\frac{83\pi^4}{45}+\frac{7052\zeta_3}{9}
                                -\frac{88\pi^2\zeta_3}{9}-272\zeta_5) \\
                     &+&C_F^2T_FN_F(\frac{5906}{27}-\frac{52\pi^2}{9}-\frac{56\pi^4}{27}+\frac{1024\zeta_3}{9})\nn \\
                     &+&C_FC_AT_FN_F(-\frac{34636}{729}+\frac{5188\pi^2}{243}+\frac{44\pi^4}{45}
                                   -\frac{3856\zeta_3}{27})\nn \\
                     &+&C_FT_F^2N_F^2(\frac{19336}{729}-\frac{80\pi^2}{27}-\frac{64\zeta_3}{27}).\nn
\eea
Finally, the coefficients of the $\beta$-function in Eq.~(\ref{eq:pertexpanomdim}), up to four loops~\cite{Tarasov:1980au,Larin:1993tp,vanRitbergen:1997va},  are given by
\bea
\beta_0 &=& \frac{11}{3} C_A - \frac{4}{3}T_F n_f, \nn \\
\beta_1 &=& \frac{34}{3}C_A^2 -\frac{20}{3}C_AT_fn_f-4C_FT_Fn_f,  \\
\beta_2 &=& \frac{2857}{54}C_A^3 + T_f n_f (2 C_F^2 -\frac{205}{9}C_F C_A-\frac{1415}{27}C_A^2) + T_f^2n_f^2(\frac{44}{9}C_F + \frac{158}{27}C_A), \nn \\
\beta_3 &=& C_A^4 \Big ( \frac{150653}{486}-\frac{44\zeta_3}{9}\Big ) + C_A^3 T_F n_f \Big (-\frac{39143}{81}+\frac{136\zeta_3}{3} \Big ) + C_A^2C_FT_Fn_f\Big(\frac{7073}{243}-\frac{656\zeta_3}{9} \Big) \nn\\
&+& C_AC_F^2T_Fn_f \Big (-\frac{4204}{27} + \frac{352\zeta_3}{9} \Big ) + 46 C_F^3T_Fn_f+ C_A^2T_F^2n_f^2\Big ( \frac{7930}{81} + \frac{224\zeta_3}{9}\Big ) \nn  \\
&+& C_F^2T_F^2n_f^2 \Big (\frac{1352}{27} -\frac{704\zeta_3}{9}\Big ) + C_AC_FT_F^2n_f^2 \Big ( \frac{17152}{243}+\frac{448\zeta_3}{9}\Big ) + \frac{424}{243}C_AT_F^3n_f^3+\frac{1232}{243}C_FT_F^3n_f^3\nn \\
&+& \frac{d_A^{abcd}d_A^{abcd}}{N_A}\Big (-\frac{80}{9}+\frac{704\zeta_3}{3} \Big ) + n_f\frac{d_F^{abcd}d_A^{abcd}}{N_A}\Big (\frac{512}{9}-\frac{1664\zeta_3}{3} \Big ) + n_f^2\frac{d_F^{abcd}d_F^{abcd}}{N_A} \Big (-\frac{704}{9}+\frac{512\zeta_3}{3} \Big ).\nn
\eea

The solution to the RG equations for the hard function in Eqs.~(\ref{eq:HRG}), (\ref{eq:Hanom})  and (\ref{eq:Canom}) has the form in Eq.~(\ref{eq:hardfuncevol}), where the hard function evolution factor has the form
\bea
\label{eq:UH}
U_H(Q^2,\mu,\mu_H) &=& \exp \Big[  4 C_FS(\mu,\mu_H)-2A_H (\mu,\mu_H)  \Big ]\Big ( \frac{\mu_H^2}{Q^2} \Big )^{2C_FA(\mu,\mu_H)}, 
\eea
where the functions $S, A,$ and $A_H$ are defined as
\bea
\label{eq:Smufmui_Amufmui_AHmufmui}
 S(\mu_f,\mu_i) &=& -\int_{\alpha_s(\mu_i)}^{\alpha_s(\mu_f)} \frac{d\alpha}{\beta[\alpha]}\gamma_{\text{cusp}}[\alpha] \int_{\alpha_s(\mu_i)}^\alpha \frac{d\alpha '}{\beta[\alpha ']},\nn \\
A(\mu_f,\mu_i) &=& -\int_{\alpha_s(\mu_i)}^{\alpha_s(\mu_f)}\frac{d\alpha}{\beta[\alpha]}\gamma_{\text{cusp}}[\alpha],  \\
A_H(\mu_f,\mu_i) &=& -\int_{\alpha_s(\mu_i)}^{\alpha_s(\mu_f)}\frac{d\alpha}{\beta[\alpha]}\gamma^{q}[\alpha], \nn
\eea
The perturbative expansion of $S(\mu_f,\mu_i)$ needed for N$^3$LL resummation is given by
\bea
\label{eq:SpertexpN3LL}
S(\mu_f,\mu_i) &=& \frac{\gamma_0^{\text{cusp}}}{4\beta_0^2}\Bigg \{ \frac{4\pi}{\alpha_s(\mu_i)} \Big ( 1-\frac{1}{r}-\ln r\Big ) + \Big ( \frac{\gamma_1^{\text{cusp}}}{\gamma_0^{\text{cusp}}}-\frac{\beta_1}{\beta_0}\Big )(1-r+\ln r )+\frac{\beta_1}{2\beta_0}\ln^2r\nn \\
&+& \frac{\alpha_s(\mu_i)}{4\pi}\Bigg [\Big ( \frac{\beta_1\gamma_1}{\beta_0\gamma_0^{\text{cusp}}} - \frac{\beta_2}{\beta_0} \Big )(1-r+r\ln r) +\Big ( \frac{\beta_1^2}{\beta_0^2}-\frac{\beta_2}{\beta_0}\Big )(1-r)\ln r  \nn\\
&-&\Big(\frac{\beta_1^2}{\beta_0^2}-\frac{\beta_2}{\beta_0}-\frac{\beta_1 \gamma_1^{\text{cusp}}}{\beta_0\gamma_0^{\text{cusp}}}+\frac{\gamma_2^{\text{cusp}}}{\gamma_0^{\text{cusp}}}\Big)\frac{(1-r)^2}{2}\Bigg ] \\
&+&\left [\frac{\alpha_s(\mu_i)}{4\pi}\right ]^2\Bigg [
                                       \left  (\frac{\beta_1\beta_2}{\beta_0^2} - \frac{\beta_1^3}{2\beta_0^3}- \frac{\beta_3}{2\beta_0} +\left (\frac{\gamma_2^{\text{cusp}}}{\gamma_0^{\text{cusp}}}-\frac{\beta_2}{\beta_0}+\frac{\beta_1^2}{\beta_0^2}-\frac{\beta_1\gamma_1^{\rm cusp}}{\beta_0\gamma_0^{\rm cusp}}\right)\frac{\beta_1r^2}{2\beta_0} \right )\ln r \nn \\
                                        &+& \left (\frac{\gamma_3^{\rm cusp}}{\gamma_0^{\rm cusp}}-\frac{\beta_3}{\beta_0} + \frac{2\beta_1\beta_2}{\beta_0^2}+\frac{\beta_1^2}{\beta_0^2}\left (\frac{\gamma_1^{\rm cusp}}{\gamma_0^{\rm cusp}} - \frac{\beta_1}{\beta_0}\right )-\frac{\beta_2\gamma_1^{\rm cusp}}{\beta_0\gamma_0^{\rm cusp}}
                                        -\frac{\beta_1\gamma_2^{\rm cusp}}{\beta_0\gamma_0^{\rm cusp}}
                                        \right )\frac{(1-r)^2}{3} \nn \\
                                        &+&\left (\frac{3\beta_3}{4\beta_0}-\frac{\gamma_3^{\rm cusp}}{2\gamma_0^{\rm cusp}}
                                       +\frac{\beta_1^3}{\beta_0^3}-\frac{3\beta_1^2\gamma_1^{\rm cusp}}
                                       {4\beta_0^2\gamma_0^{\rm cusp}}
                                        +\frac{\beta_2\gamma_1^{\rm cusp}}{\beta_0\gamma_0^{\rm cusp}}
                                        +\frac{\beta_1\gamma_2^{\rm cusp}}{4\beta_0\gamma_0^{\rm cusp}}
                                        -\frac{7 \beta_1\beta_2}{4\beta_0^2}\right )(1-r)^2\nn \\
                                        &+&\left (\frac{\beta_1\beta_2}{\beta_0^2}-\frac{\beta_3}{\beta_0}
                                        -\frac{\beta_1^2\gamma_1^{\rm cusp}}{\beta_0^2\gamma_0^{\rm cusp}}
                                        +\frac{\beta_1\gamma_2^{\rm cusp}}{\beta_0\gamma_0^{\rm cusp}}
                                         \right)\frac{1-r}{2} \nn
                                      \Bigg ]
\Bigg \},  
\eea
the corresponding perturbative expansion for $A(\mu_f,\mu_i)$ is given by
\bea
\label{Aevo}
A(\mu_f,\mu_i)&=& \frac{\gamma_0^{\text{cusp}}}{2\beta_0}\Bigg\{
\log r + \,
\frac{\alpha_s(\mu_i)}{4\pi}\left(\frac{\gamma_1^{\text{cusp}}}{\gamma_0^{\text{cusp}}}-\frac{\beta_1}{\beta_0} \right)\nn \\
&&
+ \left[\frac{\alpha_s(\mu_i)}{4\pi}\right]^2
\Bigg[\frac{\gamma_2^{\text{cusp}}}{\gamma_0^{\text{cusp}}}-\frac{\beta_2}{\beta_0}
-\frac{\beta_1}{\beta_0}\,
\left(\frac{\gamma_1^{\text{cusp}}}{\gamma_0^{\text{cusp}}}-\frac{\beta_1}{\beta_0} \right)
\Bigg]\frac{r^2-1}{2}\nn \\
&&
+ \frac{1}{3}\left[\frac{\alpha_s(\mu_i)}{4\pi}\right]^3\Bigg [\frac{\gamma_3^{\rm cusp}}{\gamma_0^{\rm cusp}}-\frac{\beta_3}{\beta_0} + \frac{\gamma_1^{\rm cusp}}{\gamma_0^{\rm cusp}}\left (\frac{\beta_1^2}{\beta_0^2}-\frac{\beta_2}{\beta_0} \right )  \\
&&
-\frac{\beta_1}{\beta_0}\left (\frac{\beta_1^2}{\beta_0^2}-\frac{2\beta_2}{\beta_0} +\frac{\gamma_2^{\rm cusp}}{\gamma_0^{\rm cusp}}\right )\Bigg ]\left (r^3-1\right )^3
\Bigg \}, \nn 
\eea
and finally, the corresponding expansion for $A_H(\mu_f,\mu_i)$ is given by
\bea
\label{AHevo}
A_H(\mu_f,\mu_i)&=& \frac{\gamma_0^{q}}{2\beta_0}\Bigg\{
\log r + \,
\frac{\alpha_s(\mu_i)}{4\pi}\left(\frac{\gamma_1^{q}}{\gamma_0^{q}}-\frac{\beta_1}{\beta_0} \right)(r-1)
 \\
&&
+ \left[\frac{\alpha_s(\mu_i)}{4\pi}\right]^2
\Bigg[\frac{\gamma_2^{q}}{\gamma_0^{q}}-\frac{\beta_2}{\beta_0}
-\frac{\beta_1}{\beta_0}\,
\left(\frac{\gamma_1^{q}}{\gamma_0^{q}}-\frac{\beta_1}{\beta_0} \right)
\Bigg]\frac{r^2-1}{2}\nn
\Bigg \}. \nn 
\eea

\subsection{Beam, jet, and soft functions}
The RG equations for the beam, jet, and soft functions in momentum are given by
\bea
\mu \frac{d}{d\mu} B_q(t,x,\mu) &=& \int dt' \> \gamma_B(t-t',\mu) \>B_q(t',x,\mu),\nn \\
\mu \frac{d}{d\mu} J(s,\mu) &=& \int ds' \> \gamma_J(s-s',\mu) J(s',\mu),\nn \\
\mu \frac{d}{d\mu}{\cal S}(\tau,\mu) &=& \int d\tau '\>\gamma_S(\tau-\tau',\mu){\cal S}(\tau',\mu),
\eea
where the anomalous dimensions have the form
\bea
\gamma_B(t,\mu) &=& -2C_F \gamma_{\text{cusp}}(\alpha_s)\frac{1}{\mu^2}{\cal L}_0(t/\mu^2)  + \gamma_B^q (\alpha_s) \delta(t), \nn \\
\gamma_J(s,\mu) &=& -2 C_F\gamma_{\text{cusp}} (\alpha_s) \>\frac{1}{\mu^2}{\cal L}_0(s/\mu^2) + \gamma^q_J(\alpha_s)\> \delta(s),  \\
\gamma_S(\tau,\mu) &=& 4C_F \gamma_{\text{cusp}}(\alpha_s)\>\frac{1}{\tilde{\mu}}{\cal L}_0(\tau/\tilde{\mu}) + 2\gamma_S^q(\alpha_s)\>\delta(\tau), \nn
\eea
where $\tilde{\mu}=\mu r_s$ and $r_s$ is defined in Eq.~(\ref{eq:tildemuS}). The corresponding RG equations for the position space for the jet, beam, and soft functions, related to their corresponding momentum space definitions as in Eq.~(\ref{eq:FT_BJS}), take the multiplicative form
\bea
\label{eq:RGJBS}
\mu \frac{d}{d\mu} J(y,\mu) &=& \gamma_J(y,\mu) J(y,\mu), \nn \\
\mu \frac{d}{d\mu}B_q(y,x,\mu)&=& \gamma_B(y,\mu) B_q(y,x,\mu), \\
\mu\frac{d}{d\mu}{\cal S}(y,\mu) &=& \gamma_S(y,\mu)\> {\cal S}(y,\mu), \nn 
\eea
and the position and mometum space anomalous dimensions are related by
\bea
\label{eq:BJSanomPos}
\gamma_B(y,\mu) &=& \int dt \> e^{-ity} \gamma_B(t,\mu), \nn \\
\gamma_J(y,\mu) &=& \int ds \> e^{-iys} \gamma_J(s,\mu), \\
\gamma_S(y,\mu) &=& \int dk \> e^{-i\tau y}\> \gamma_S(\tau,\mu). \nn 
\eea
Using Eqs.~(\ref{eq:BJSanomPos}) and (\ref{eq:LnFT}), the position space anomalous dimensions have the form
\bea
\gamma_B(y,\mu) &=& 2C_F \gamma_{\text{cusp}}(\alpha_s) \ln(i y \mu^2 e^{\gamma_E}) + \gamma_B^q(\alpha_s), \nn \\
\gamma_J(y,\mu) &=& 2C_F \gamma_{\text{cusp}}(\alpha_s) \ln(i y \mu^2 e^{\gamma_E}) + \gamma_J^q(\alpha_s),  \\
\gamma_S(y,\mu)&=&-4C_F \gamma_{\text{cusp}}(\alpha_s) \ln(i y \mu e^{\gamma_E}) + 2\gamma_S^q(\alpha_s).  \nn
\eea
Solving the RG equations in Eq.~(\ref{eq:RGJBS}) gives the beam, jet, and soft functions evolved to any arbirary scale, $\mu$, from their values at their natural scales $\mu_B, \mu_J,$ and $\mu_S$, respectively, where large logarithms in their perturbative expansions are minimized. They have the general form given in Eq.~(\ref{eq:BJSevolPos}), 
where the $U_B,U_J,$ and $U_S$ denote the RG evolution factors have the form
\bea
\label{eq:UBJS}
U_B(y_B,\mu_f,\mu_i) &=& \text{exp}\Big [ -4 C_F S(\mu_f,\mu_i) - A_B(\mu_f,\mu_i)\Big ] \Big ( i y_B \mu_i^2 e^{\gamma_E}\Big )^{-2C_F A(\mu_f,\mu_i)},\nn \\
U_J(y_J,\mu_f,\mu_i) &=& \text{exp}\Big [ -4 C_F S(\mu_f,\mu_i) - A_J(\mu_f,\mu_i)\Big ] \Big ( i y_J \mu_i^2 e^{\gamma_E}\Big )^{-2C_F A(\mu_f,\mu_i)},  \\
U_S(y_S,\mu_f,\mu_i) &=& \text{exp}\Big [ 4C_FS(\mu_f,\mu_i)- 2 A_S(\mu_f,\mu_i)\Big] \Big [ (i y_S \mu_i r_S e^{\gamma_E})^2\Big ]^{2C_FA(\mu_f,\mu_i)}, \nn
\eea
where the function $S$ is defined in Eq.~(\ref{eq:Smufmui_Amufmui_AHmufmui}) and its perturbative expansion needed for N$^3$LL resummation is given in Eq.~(\ref{eq:SpertexpN3LL}). The functions $A_B, A_J,$ and $A_S$ are defined as
\bea
\label{eq:ABAJAS}
A_B(\mu_f,\mu_i)&=& -\int_{\alpha_s(\mu_i)}^{\alpha_s(\mu_f)}\frac{d\alpha}{\beta[\alpha]}\gamma^{q}_B[\alpha],\nn \\
A_J(\mu_f,\mu_i)&=&= -\int_{\alpha_s(\mu_i)}^{\alpha_s(\mu_f)}\frac{d\alpha}{\beta[\alpha]}\gamma^{q}_J[\alpha],\nn \\
A_S(\mu,\mu_S) &=& -\int_{\alpha_s(\mu_i)}^{\alpha_s(\mu_f)}\frac{d\alpha}{\beta[\alpha]}\gamma_S^q[\alpha],
\eea
where $\gamma^{q}_B, \gamma^{q}_J,\gamma^{q}_S$, and $\gamma^{q}$ satisfy the relations
\bea
\label{eq:relgamBJSq}
\gamma^q_B &=& \gamma^q_J, \qquad \gamma^q_S = - \gamma^q_J -\gamma^q_B-\gamma^q,
\eea
where the second relation reflects from the cancellation of renomalization scale dependence between the hard, beam, jet, and soft functions in the resummation cross section formula of Eq.~(\ref{eq:factorization_resum_FT}). The perturbative expansion of $\gamma^q_J$ up to three loops~\cite{Becher:2006mr}, needed for N$^3$LL resummation is given by
\bea
\label{eq:gamqJpertexp}
\gamma^{q}_{J_0} &=& 6 C_F, \nn \\
\gamma^{q}_{J_1} &=& C_F \Big [(\frac{146}{9}-80\zeta_3)C_A +(3-4\pi^2+ 48\zeta_3) C_F +(\frac{121}{9} +\frac{2\pi^2}{3})\beta_0 \Big ],  \\
\gamma^{q}_{J_2} &=& -2 C_F^3\left (-\frac{29}{2} - 3\pi^2 - \frac{8\pi^4}{5}-68\zeta_3+\frac{16\pi^2\zeta_3}{3} +240\zeta_5\right )\nn \\
&-& 2 C_F^2 C_A\left (-\frac{151}{4} + \frac{205\pi^2}{9} +\frac{247\pi^4}{135} -\frac{844\zeta_3}{3} -\frac{8\pi^2\zeta_3}{3}-120\zeta_5\right ) \nn \\
&-&2C_F C_A^2 \left (-\frac{412907}{2916}-\frac{419\pi^2}{243}-\frac{19\pi^4}{10}+\frac{5500\zeta_3}{9}-\frac{88\pi^2\zeta_3}{9}-232\zeta_5 \right )\nn \\
&-&2C_F^2T_Fn_f\left (\frac{4664}{27}-\frac{32\pi^2}{9}-\frac{164\pi^4}{135}+\frac{208\zeta_3}{9} \right ) \nn \\
&-& 2C_F C_A T_F n_f\left (-\frac{5476}{729} + \frac{1180\pi^2}{243} + \frac{46\pi^4}{45}-\frac{2656\zeta_3}{27}\right ) \nn \\
&-& 2 C_F T_F^2 n_f^2 \left (\frac{13828}{729}-\frac{80\pi^2}{81}-\frac{256\zeta_3}{27} \right ), \nn 
\eea
and the corresponding expressions for $\gamma^q_B$ and $\gamma^q_S$ can be obtained from a combination of Eqs.~(\ref{eq:relgamBJSq}), (\ref{eq:gamqJpertexp}), and (\ref{eq:gamqpertexpN3LL}). The corresponding expressions for the perturbative expansions of $A_B, A_J,$ and $A_S$ in Eq.~(\ref{eq:ABAJAS}) needed for N$^3$LL resummation can be obtained by replacing $\gamma^{q}_{0,1,2}\to $ with $\gamma^{q}_{B_{0,1,2}}, \gamma^{q}_{J_{0,1,2}},$ and $\gamma^{q}_{S_{0,1,2}}$, respectively, in Eq.~(\ref{AHevo}).

\subsection{Product of RG Evolution Factors}
The factorization and resummation formula in Eq.~(\ref{eq:factorization_resum_FT}), contains a product of the position space RG evolution factors for the hard, beam, jet, and soft functions, given by
\bea
\label{eq:Utotal}
U_{\text{total}} &\equiv& U_H(\xi^2,\mu,\mu_H)U_B(\frac{y_\tau}{Q_a},\mu,\mu_B)U_J(\frac{y_\tau}{Q_J},\mu,\mu_J)U_S(y_\tau,\mu,\mu_S).
\eea
Using the results for the corresponding RG evolution factors in Eqs.~(\ref{eq:UH}) and (\ref{eq:UBJS}), the combined evolution factor, $U_{\rm total}$, is given by
\bea
\label{Utotal}
U_{\text{total}}  = U_{HBJS}\> (r_S)^{4C_F A(\mu,\mu_S)}  (r_B)^{-2C_F A(\mu,\mu_B)}
 (r_J)^{-2C_F A(\mu,\mu_J)}  (iy_\tau \mu_S e^{\gamma_E})^{\omega}, 
\eea
where we have defined, $U_{HBJS}$, as
\bea
\label{UHBJS}
U_{\text{HBJS}} &\equiv& \text{exp}\Big [4C_F [S(\mu,\mu_H)+S(\mu,\mu_S)-S(\mu,\mu_B)-S(\mu,\mu_J)] \Big ]\nn \\
&\times&\text{exp}\Big [-2A_H(\mu,\mu_H)-A_B(\mu,\mu_B)-A_J(\mu,\mu_J)-2A_S(\mu,\mu_S) \Big ]  \\
&\times&\>\Big ( \frac{\mu_H^2}{\xi^2} \Big )^{2C_FA(\mu,\mu_H)}, \nn
\eea
and, $\omega=\omega(\mu,\mu_B,\mu_J,\mu_S)$, as
\bea
\label{omega}
\omega(\mu,\mu_B,\mu_J,\mu_S)&=& 2C_F\big [ 2A(\mu,\mu_S)-A(\mu,\mu_B)-A(\mu,\mu_J)\big ],\nn \\
&=& 2C_F\big [ A(\mu_B,\mu_S) + A(\mu_J,\mu_S)\big ].
\eea


 \section{Numerical Implementation of Resummation Factorization Formula}
 \label{appex:NumFacFormula}

In order to implement code that can generate numerical results in the resummation region, it is useful work with the position space resummation factorization formula in Eq.~(\ref{eq:factorization_resum_FT}). Using Eq.~(\ref{eq:Fmod_FT}) to rewrite the model soft function in momentum space, and using Eq.~(\ref{Utotal}), the resummation factorization formula in Eq.~(\ref{eq:factorization_resum_FT}) can be brought into the form
\bea
\label{eq:factorization_resum_numerical}
d\sigma_{\rm resum} \left [\tau_1,P_{J_T},y_J \right ] &=&\sigma_0 \>U_{HBJS}( \hat{s}_{aJ} )^{2C_FA(\mu,\mu_S)} (r_B)^{-2C_FA(\mu,\mu_B)} (r_J)^{-2C_FA(\mu,\mu_J)} \nn \\
&&\times  H(\xi^2, \mu_H)   \int du \int \frac{dy_\tau}{2\pi}   e^{iy_\tau (\tau_1-u)}\big (i y_\tau\mu_S e^{\gamma_E} \big )^{\omega} \nn \\
&&\times J(\frac{y_\tau}{Q_J}, \mu_J)\>{\cal S}_{\text{part}.}\left(y_\tau,\mu_S\right)\>F_{\text{mod}.}\left(u \right),\nn \\
&&\times \Big [ \sum_{q}\sum_{i}  L_q \int_{x_*}^1 \frac{dz}{z} \> {\cal I}_{qi}\left(\frac{x_*}{z}, \frac{y_\tau}{Q_a},  \mu_B\right) f_{i/p}(z,\mu_B) \nn \\
&&+  \sum_{\bar{q}}\sum_{i}  L_{\bar{q}} \int_{x_*}^1 \frac{dz}{z} \> {\cal I}_{\bar{q}i}\left( \frac{x_*}{z},  \frac{y_\tau}{Q_a}, \mu_B\right) f_{i/p}(z,\mu_B) \Big ].
\eea
The hard function has a perturbative expansion expressed as in Eq.~(\ref{eq:Hexp}). The perturbative expansions of the soft, jet, and beam functions in Eqs.~(\ref{eq:Sexp}), (\ref{eq:Jexp}), and (\ref{eq:Iijexpy}) can be re-expressed in powers of $L_S = \ln (i y_{\tau} \mu_S e^{\gamma_E})$, when $y_J=y_\tau/Q_J$ and $y_B=y_\tau/Q_B$, as
\bea
\label{eq:pertexp}
S(y_\tau,\mu_S)_{\rm part.} &=& \sum_{n=0}^\infty \sum_{m=0}^{2n} \left [\frac{\alpha_s(\mu_S)}{4\pi} \right ]^n L_S^m  \>S^{(n)}_m, \nn \\
J_{i}(y_\tau/Q_J,\mu_J) &=& \sum_{n=0}^\infty \sum_{m=0}^{2n} \left [\frac{\alpha_s(\mu_J)}{4\pi} \right ]^n L_S^m  \>J^{(n)}_m, \nn \\
{\cal I}_{(q,\bar{q})i}^{(n)}(y_\tau/Q_B,x,\mu_B) &=& \sum_{m=0}^{2n} \sum_{k=-2}^\infty \>   \left [\frac{\alpha_s(\mu_B)}{4\pi} \right ]^n \> L_S^m\> {\cal L}_k(1-x)\> {\cal I}_{(q,\bar{q})i,mk}^{(n)}(x) ,
\eea
where, for ease of notation, we have defined $
{\cal L}_{-2}(x) \equiv \delta(x)$ and $
{\cal L}_{-1}(x) \equiv \theta(x)
$. Here, $S^{(n)}_m$ and $ J^{(n)}_m,$  denote the coefficients of $\left (\alpha_s(\mu_S)/\left (4\pi\right) \right )^n L_S^m $ in the soft and jet function perturbative series, respectively. $ {\cal I}_{(q,\bar{q})i,mk}^{(n)}(x)$ denotes the coefficient of ${\cal L}_k(1-x) \left (\alpha_s(\mu_S)/\left (4\pi\right) \right )^n L_S^m $ in the perturbative series for the beam function. In arriving at this form of the perturbative series for the soft, jet, and beam functions, we defined two new variables
\bea
r_B = \frac{\mu_B^2}{Q_B \mu_S}, \qquad 
 r_J = \frac{\mu_J^2}{Q_J \mu_S}, 
 \eea
which along with the definition of $r_S$ in Eq.~(\ref{eq:tildemuS}) allowed us to write the logarithms $\tilde{L}_S$, $L_J$, $L_B$ that appear in the position space perturbative expansions in Eqs.~(\ref{eq:Sexp}), (\ref{eq:Jexp}), and (\ref{eq:Iijexpy}), for the soft, jet, and beam functions, respectively, as
\bea
\label{eq:LSLJLB}
\tilde{L}_S &=& L_S + \ln r_S, \nn \\
L_J &=& \ln (iy_J \mu_J^2e^{\gamma_E}) = L_S + \ln r_J,  \\
L_B &=& \ln(iy_B\mu_B^2e^{\gamma_E}) = L_S + \ln r_B , \nn
\eea
when $y_J=y_\tau/Q_J$ and $y_B=y_\tau/Q_B$.

The hard function coefficient functions $H^{(n)}$ in Eq.~(\ref{eq:pertexp}) are functions of $\ln(\xi^2/\mu_H^2)$ and the coefficients up to ${\cal O}(\alpha_s^2)$~\cite{Manohar:2003vb,Idilbi_2006,Becher:2006mr}, $H^{(0,1,2)}$, are given in Eq.~(\ref{H(n)}).  The partonic soft function~\cite{Jouttenus:2011wh,Boughezal:2015eha}, jet function~\cite{Bosch:2004th,Becher:2006qw,Becher:2010pd}, and beam function~\cite{Stewart:2009yx, Stewart:2010qs, Mantry:2009qz,Berger:2010xi,Gaunt:2014xga,Gaunt:2014cfa}  are also known up to ${\cal O}(\alpha_s^2)$. These fixed order results are used to extract the coefficient functions $S^{(n)}_m, J^{(n)}_m,$ and ${\cal I}_{ij,mk}^{(n)}(x)$  up to ${\cal O}(\alpha_s^2)$. Using these fixed order expansions of the hard, beam, jet, and partonic soft functions in Eq.~(\ref{eq:pertexp}), the resummation formula can be brought into the final form
\bea
\label{eq:fac_resum_numeric}
d\sigma_{\rm resum} \left [\tau_1,P_{J_T},y_J \right ] &=&\sigma_0 \>U_{HBJS}( r_S )^{4C_FA(\mu,\mu_S)} (r_B)^{-2C_FA(\mu,\mu_B)} (r_J)^{-2C_FA(\mu,\mu_J)} \nn \\
&\times& \sum_{\substack{n_1,n_2,\\n_3,n_4}} \sum_{\substack{m_2,m_3,\\m_4}}
\left [\frac{\alpha_s(\mu_H)}{4\pi}\right]^{n_1} \left [\frac{\alpha_s(\mu_J)}{4\pi}\right]^{n_2}\left [\frac{\alpha_s(\mu_B)}{4\pi}\right]^{n_3} \left [\frac{\alpha_s(\mu_S)}{4\pi}\right]^{n_4}  \\
&\times& H^{(n_1)}\>J^{(n_2)}_{m_2}\> S^{(n_4)}_{m_4} \> {\cal K}^{(n_3)}_{m_3} \int du \>F_{\text{mod}.}\left(u \right) d_{m_2+m_3+m_4}(\tau_1 -u, \omega, \mu_S) , \nn
\eea
where we have defined the coefficient functions ${\cal K}^{(n)}_{m}$ as
\bea
{\cal K}^{(n)}_{m} &\equiv&  \sum_{q} \sum_{i} L_q \int_{x_*}^1 \frac{dz}{z} \>\sum_{k} {\cal I}^{(n)}_{qi,m k}\left(z\right) {\cal L}_k(1-z)f_{i}(\frac{x_*}{z},\mu_B) \nn \\
&+& \sum_{\bar{q}} \sum_{i} L_{\bar{q}} \int_{x_*}^1 \frac{dz}{z} \>\sum_{k} {\cal I}^{(n)}_{\bar{q}i,m k}\left(z\right) {\cal L}_k(1-z)f_{i}(\frac{x_*}{z},\mu_B), 
\eea
and the $d_m(\tau_1,\omega,\mu_S)$ functions as
\bea
\label{dm}
d_m(\tau_1, \omega, \mu_S) \equiv \int \frac{dy_\tau}{2\pi}   e^{iy_\tau \tau_1}\big (i y_\tau\mu_S e^{\gamma_E} \big )^{\omega} L_S^{m} = \partial_\omega^{m} \Bigg \{ \frac{e^{\omega \gamma_E}}{\Gamma(-\omega)} \frac{1}{\mu_S} \> \left [ \frac{\mu_S^{1+\omega}\theta(\tau_1)}{(\tau_1)^{1+\omega}} \right ]_+\Bigg \}.
\eea
In arriving at this result we made use of the relation $  (iy_\tau \mu_S e^{\gamma_E})^\omega L_S^n = \partial_\omega^n(iy_\tau \mu_S e^{\gamma_E})^\omega$ and the identity in Eq.~(\ref{eq:PlusDistIdenitity}). In the factorization formula, we always have $\omega = \omega(\mu_B,\mu_J,\mu_S) < 0$ as seen from Eq.~(\ref{omega}) and the hierarchy of scales $\mu_S < \mu_B, \mu_J$. This allows us to drop the plus-prescription in the numerical evaluation of the $d_m(\tau_1,\omega,\mu_S)$ functions. The explicit results for $d_{0,1,2,3,4}$ are: 
\bea
\label{derv}
d_0(\omega,\tau_1-u,\mu_S)&=& \frac{e^{\omega \gamma_E}}{\Gamma(-\omega)}\frac{1}{\mu_S}\Big (\frac{\mu_S}{\tau_1-u}\Big )^{1+\omega}
\nn \\
d_1(\omega,\tau_1-u,\mu_S)&=& \frac{e^{\omega \gamma_E}}{\Gamma(-\omega)}\frac{1}{\mu_S}\Big (\frac{\mu_S}{\tau_1-u}\Big )^{1+\omega}\Big [ \ln \frac{\mu_Se^{\gamma_E}}{\tau_1-u} +\psi^{(0)}(-\omega)\Big ] ,\nn \\
d_2(\omega,\tau_1-u,\mu_S)&=& \frac{e^{\omega \gamma_E}}{\Gamma(-\omega)}\frac{1}{\mu_S}\Big (\frac{\mu_S }{\tau_1-u}\Big )^{1+\omega}\Big [\Big ( \ln \frac{\mu_S e^{\gamma_E}}{\tau_1-u} +\psi^{(0)}(-\omega)\Big )^2
-\psi^{(1)}(-\omega)\Big ].\nn \\
d_3(\omega,\tau_1-u,\mu_S)&=& \frac{e^{\omega \gamma_E}}{\Gamma(-\omega)}\frac{1}{\mu_S}\Big (\frac{\mu_S }{\tau_1-u}\Big )^{1+\omega}\Big [\Big ( \ln \frac{\mu_S e^{\gamma_E}}{\tau_1-u} +\psi_0(-\omega)\Big )^3\nn \\
&-&3\psi^{(1)}(-\omega)\Big ( \ln \frac{\mu_S e^{\gamma_E}}{\tau_1-u} +\psi^{(0)}(-\omega)\Big ) + \psi^{(2)}(-\omega)\Big ],\nn \\
d_4(\omega,\tau_1-u,\mu_S)&=& \frac{e^{\omega \gamma_E}}{\Gamma(-\omega)}\frac{1}{\mu_S}\Big (\frac{\mu_S }{\tau_1-u}\Big )^{1+\omega}
\Big [ \left [\psi^{(0)}(-\omega)\right ]^4 + 3\left [\psi^{(1)}(-\omega)\right ]^2 -\psi^{(3)}(-\omega) \nn \\
 &+&
 \ln^4\left(\frac{\mu_S e^{\gamma_E}}{\tau_1-u}\right)
 + 4\left [\psi^{(0)}(-\omega)\right ]^3 \ln\frac{\mu_S e^{\gamma_E}}{\tau_1-u}\nn \\
 &+& 6\left [\psi^{(0)}(-\omega)\right ]^2 \left (  \ln^2\left(\frac{\mu_S e^{\gamma_E}}{\tau_1-u}\right) - \psi^{(1)}(-\omega)\right )  \\
 &-& 6\psi^{(1)}(-\omega) \ln^2 \left(\frac{\mu_S e^{\gamma_E}}{\tau_1-u}\right ) +  4\psi^{(2)}(-\omega)\ln\frac{\mu_S e^{\gamma_E}}{\tau_1-u}  \nn \\
 &+& 4\psi^{(0)}(-\omega)\left ( \ln^3\left(\frac{\mu_S e^{\gamma_E}}{\tau_1-u}\right) - 3 \psi^{(1)}(-\omega)\ln\frac{\mu_S e^{\gamma_E}}{\tau_1-u} +\psi^{(2)}(-\omega)\right )
\Big ],\nn 
\eea
where $\psi^{(n)}(x)$ is the PolyGamma function of order $n$
\bea
\psi^{(n)}(x) = \frac{d^{n+1}}{dx^{n+1}} \ln \Gamma(x).
\eea
Eq.~(\ref{eq:fac_resum_numeric}) serves as the master formula for the numerical implementation of the resummed factorization formula.

 
 \section{Non-Perturbative Shape Function}
 \label{appex:NPsoft function}

In general, the 1-jettiness shape function  $F_{\rm mod}(u)$, appearing in Eq.~(\ref{eq:soft-model-conv}),  can depend on the null beam and jet reference vectors $n_B^\mu =(1,\vec{n}_B)$ and $n_J^\mu =(1,\vec{n}_J)$, respectively. Here $\vec{n}_B$ and $\vec{n}_J$ are unit 3-vectors that point along the beam and leading jet directions, respectively. Note that for each event, the 3-vector $\vec{n}_J$ can point in a different direction, corresponding to the leading jet. In this section, we derive an analytic formula that explicitly shows how one can incorporate the dynamical dependence on the beam and jet reference vectors into a model for the shape function, $F_{\rm mod}(u)$. We derive this result by following and building on the analysis in Ref.~\cite{Kang:2013nha}.

The soft function that appears in the factorization formula has the form
\bea
\label{softtau1}
{\cal S}(\tau_1,\mu) = \int \>dk_B \> \int dk_J\> \delta(\tau_1 - k_B - k_J) \>{\cal S}(k_B, k_J, \mu),
\eea
as in Eq.~(\ref{eq:soft-projection}), where ${\cal S}(k_B, k_J, \mu)$ is the generalized hemisphere soft function where the arguments $k_B,k_J$, correspond to the contribution to $\tau_1$ of soft radiation grouped with the beam and jet reference vector directions, respectively. The field theoretic definition of the generalized soft function is~\cite{Kang:2013wca, Kang:2013nha}:
\bea
\label{genhemisoft}
{\cal S}(k_B, k_J, \mu) &=& \frac{1}{N_c} \>{\rm tr}\>\sum_{X_s} \Big |\langle X_s | [Y_{n_J}^\dagger Y_{n_B}] (0) |0\rangle \Big |^2 \nn \\
&& \delta \left [k_B - \sum_{i \in X_s} \theta \left ( \frac{q_J}{Q_J}\cdot k_i - \frac{q_B}{Q_B}\cdot k_i \right ) \frac{q_B }{Q_B} \cdot k_i\right ] \nn \\
&& \delta \left [k_J - \sum_{i \in X_s} \theta \left (  \frac{q_B}{Q_B}\cdot k_i - \frac{q_J}{Q_J}\cdot k_i \right ) \frac{q_J}{Q_J}\cdot k_i\right ].
\eea
The beam and jet reference vectors that appear in the definition of $\tau_1$, in the resummation region,  can be written as:
\bea
\label{eq:beam-jet-ref}
q_B^\mu &=& \omega_B \frac{n_B^\mu}{2}, \qquad q_J^\mu = \omega_J \frac{n_J^\mu}{2}, \nn \\
Q_B &=&\omega_B, \qquad Q_J = \omega_J.
\eea
In our work, for $\tau_1$, we have $\omega_B = x \sqrt{s}$ and  $\omega_J= 2P_{J_T}\cosh y_J$, corresponding to Eqs.~(\ref{eq:beam_ref_choices}), (\ref{eq:jet_ref_choices}), and (\ref{eq:qJ_resum}).  In general, the scalar dot product $n_B\cdot n_J$ will depend on the direction of the leading jet in each event.  

As explained in Ref.~\cite{Kang:2013nha}, around their Eq.~(133), one can define new null reference vectors, $n_B'$ and $n_J'$ as:
\bea
\label{nBnJprime}
n_B' = n_B/R_B, \qquad n_J'=n_J/R_J,
\eea
where $R_B$ and $R_J$ are defined as:
\bea
\label{RBRJ}
R_B = \sqrt{\frac{\omega_J}{Q_J}\frac{Q_B}{\omega_B}\frac{n_B\cdot n_J}{2}}, \qquad R_J = \sqrt{\frac{\omega_B}{Q_B}\frac{Q_J}{\omega_J}\frac{n_B\cdot n_J}{2}},
\eea
which leads to the result:
\bea
n_B'\cdot n_J' =2 = \>{\rm constant},
\eea
for every event.  Using Eqs.~(\ref{eq:beam-jet-ref}) and the invariance~\cite{Lee:2006nr} of the Wilson lines under  the transformations in Eq.~(\ref{nBnJprime}), generalized hemisphere soft function can be written as: 
\bea
{\cal S}(k_B, k_J, \mu) &=& \frac{1}{N_c} \>{\rm tr}\>\sum_{X_s} \Big |\langle X_s | [Y_{n_J'}^\dagger Y_{n_B'}] (0) |0\rangle \Big |^2 \nn \\
&& \frac{1}{R_B}\delta \left [\frac{k_B}{R_B} - \sum_{i \in X_s} \theta \left ( n_J'\cdot k_i - n_B'\cdot k_i \right ) n_B' \cdot k_i\right ] \nn \\
&& \frac{1}{R_J}\delta \left [\frac{k_J}{R_J}  - \sum_{i \in X_s} \theta \left ( n_B'\cdot k_i - n_J'\cdot k_i \right ) n_J' \cdot k_i\right ] ,
\eea
which corresponds to the relation~\cite{Kang:2013nha}:
\bea
\label{gen-std-hem-rel}
{\cal S}(k_B, k_J, \mu)  &=&\frac{1}{R_B R_J} {\cal S}_{\rm hemi.}(\frac{k_B}{R_B}, \frac{k_J}{R_J}, \mu). 
\eea
i.e. the generalized hemisphere soft function, ${\cal S}(k_B, k_J, \mu)$,  is related to the standard hemisphere soft function ${\cal S}_{\rm hemi.}(k_1, k_2, \mu)$, evaluated with the reference vectors $n_1$ and $n_2$ such that $n_1\cdot n_2=2$, a constant, if one makes the subsitution $k_1=k_B/R_B$ and $k_2=k_J/R_J$. Thus, all the dependence on the reference vectors $n_B$ and $n_J$ in the 1-jettiness soft function is accounted for through the factors $R_B$ and $R_J$ as shown above.

Thus, the 1-jettiness soft function in Eq.~(\ref{softtau1}) can now be written in terms of the standard hemisphere soft function as:
\bea
{\cal S}(\tau_1,\mu) = \int \>dk_B \> \int dk_J\> \delta(\tau_1 - k_B - k_J) \frac{1}{R_B R_J}{\cal S}_{\rm hemi.}(\frac{k_B}{R_B}, \frac{k_J}{R_J}, \mu).
\eea
Through a simple change of integration variables, this can be brought into the equivalent form:
\bea
\label{softproj}
{\cal S}(\tau_1,\mu) = \int \>dk_B \> \int dk_J\> \delta(\tau_1 - R_B k_B - R_J k_J) \> {\cal S}_{\rm hemi.}(k_B, k_J, \mu).  
\eea
Thus, a shape function model for ${\cal S}(\tau_1,\mu)$ can now be incorporated in terms of a shape function model for the hemisphere soft function through the convolution:
\bea
\label{shapehemi}
{\cal S}_{\rm hemi.}(k_B, k_J, \mu) = \int dk_B' \int dk_J' \> {\cal S}_{\rm hemi.}^{\rm part.}(k_B - k_B', k_J-k_J', \mu) \> {\cal S}_{\rm hemi.}^{\rm mod.}(k_B', k_J'),
\eea
where ${\cal S}_{\rm hemi.}^{\rm part.}$ is the partonic hemisphere soft function and ${\cal S}_{\rm hemi.}^{\rm mod.}$ is the model hemisphere shape function, which satisfies the normalization condition:
\bea
\int \>dk_B \> \int dk_J\> {\cal S}_{\rm hemi.}^{\rm mod.}(k_B, k_J)= 1.
\eea

Using Eq.~(\ref{shapehemi}) in Eq.~(\ref{softproj}),  following the procedure outlined in pages  17, 18, and 19 of  Ref.~\cite{Kang:2013wca}, one can  compare the result to Eq.~(\ref{eq:soft-model-conv}), to extract $F^{\rm mod.}(u)$ as:
\bea
\label{Fmod-beam-jet-ref-dependence}
F^{\rm mod.}(u) &=& \frac{1}{2}\int_{-u}^{u} d\zeta \>\> {\cal S}^{\rm mod.}(\frac{u+\zeta}{2}, \frac{u-\zeta}{2}) \nn \\
&=& \frac{1}{R_B+R_J}\int_{-u/R_J}^{u/R_B} d\zeta \>\> {\cal S}_{\rm hemi.}^{\rm mod.}(\frac{u+R_J\zeta}{R_B+R_J}, \frac{u-R_B \zeta}{R_B+R_J}),
\eea
where we have expressed $F^{\rm mod.}(u)$ in terms of the generalized hemisphere function, ${\cal S}^{\rm mod.}$, and the standard hemisphere soft function, ${\cal S}^{\rm mod.}$, in the first and second equalities, respectively.  In the first equality, the original $k_B$ and $k_J$ variables are related to the  transformed variables as $u= k_B + k_J$ and $\zeta = k_B -k_J$. In the second equality, the original $k_B$ and $k_J$ variables are related to the  transformed variables as $u=R_B k_B + R_J k_J$ and $\zeta = k_B -k_J$.
This shows the explicit dependence of the shape function $F^{\rm mod.}(u)$ on the reference vectors $n_B$ and $n_J$ through  $R_B$ and $R_J$, defined in Eq.~(\ref{RBRJ}) above.  For $\tau_1$, using Eqs.~(\ref{eq:beam-jet-ref}) and (\ref{RBRJ}), we can have:
\bea
\label{RBRJrS}
R_B=R_J= \sqrt{\frac{n_B\cdot n_J}{2}} = \sqrt{\frac{2 q_B\cdot q_J}{Q_B Q_J}} = r_S,
\eea
where $r_S$ is defined in Eq.~(\ref{eq:tildemuS}). Thus, Eq.~(\ref{Fmod-beam-jet-ref-dependence}) can be written as entirely in terms of $r_S$ as:
\bea
\label{Fmod-beam-jet-ref-dependence-2}
F^{\rm mod.}(u) 
&=& \frac{1}{2 r_S}\int_{-u/r_S}^{u/r_S} d\zeta \>\> {\cal S}_{\rm hemi.}^{\rm mod.}(\frac{u+r_S \>\zeta}{2r_S}, \frac{u-r_S \> \zeta}{2r_S})\> .
\eea
Thus, one can construct a model for the standard hemisphere shape function, ${\cal S}_{\rm hemi.}^{\rm mod.}(k_1,k_2)$ and then use it in the above equation to obtain the corresponding model for the shape function $F^{\rm mod.}(u)$ with the full dependence on the beam and jet reference vectors, encoded in $r_S$. This result that determines the shape function, $F^{\rm mod.}(u)$, for the $\tau_1$ observable in terms of the standard hemisphere function ${\cal S}_{\rm hemi.}^{\rm mod.}(k_1,k_2)$ corresponds to a degree of universality among shape functions in DIS event shapes.

Finally, we note that the first moment of the shape function $F^{\rm mod.}(u)$ can be expressed in terms of the generalized hemisphere function and the standard hemisphere function moments as
\bea
 \int du \> u\>F^{\rm mod.}(u) 
&=& \int dk_B \int dk_J \> (k_B + k_J)\> S^{\rm mod.}(k_B, k_J), \nn \\
&=& r_S \int dk_1 \int dk_2 \>(k_1 + k_2) \> S^{\rm mod.}_{\rm hemi.}(k_1, k_2),
\eea
again showing universality up to the overall factor of $r_S$. This is consistent with the expected universality~\cite{Lee:2006nr, Mateu:2012nk, Kang:2013nha} of the leading power correction, determined by the first moment of the shape function, in the tail region of the $\tau_1$-distribution.



\bibliographystyle{h-physrev3.bst}
\bibliography{disjettiness}

\begin{thebibliography}{10}

\bibitem{AbdulKhalek:2021gbh}
R.~Abdul~Khalek {\em et~al.},
\newblock (2021), 2103.05419.

\bibitem{DOE_LRP}
G.~Dodge {\em et~al.},
\newblock {\em {A New Era of Discovery: The 2023 Long Range Plan for Nuclear
  Science}}, \url{https://nuclearsciencefuture.org/}.

\bibitem{DPAPreport2022}
E.~D. P.~A. Panel,
\newblock (2022).

\bibitem{Antonelli:1999kx}
V.~Antonelli, M.~Dasgupta, and G.~P. Salam,
\newblock JHEP {\bf 0002}, 001 (2000), hep-ph/9912488.

\bibitem{Dasgupta:2001sh}
M.~Dasgupta and G.~Salam,
\newblock Phys.Lett. {\bf B512}, 323 (2001), hep-ph/0104277.

\bibitem{Dasgupta:2001eq}
M.~Dasgupta and G.~Salam,
\newblock Eur.Phys.J. {\bf C24}, 213 (2002), hep-ph/0110213.

\bibitem{Dasgupta:2002bw}
M.~Dasgupta and G.~P. Salam,
\newblock JHEP {\bf 0203}, 017 (2002), hep-ph/0203009.

\bibitem{Catani:1996vz}
S.~Catani and M.~Seymour,
\newblock Nucl.Phys. {\bf B485}, 291 (1997), hep-ph/9605323.

\bibitem{Graudenz:1997gv}
D.~Graudenz,
\newblock (1997), hep-ph/9710244.

\bibitem{Adloff:1997gq}
H1 Collaboration, C.~Adloff {\em et~al.},
\newblock Phys.Lett. {\bf B406}, 256 (1997), hep-ex/9706002.

\bibitem{Aktas:2005tz}
H1 Collaboration, A.~Aktas {\em et~al.},
\newblock Eur.Phys.J. {\bf C46}, 343 (2006), hep-ex/0512014.

\bibitem{Adloff:1999gn}
H1 Collaboration, C.~Adloff {\em et~al.},
\newblock Eur.Phys.J. {\bf C14}, 255 (2000), hep-ex/9912052.

\bibitem{Breitweg:1997ug}
ZEUS Collaboration, J.~Breitweg {\em et~al.},
\newblock Phys.Lett. {\bf B421}, 368 (1998), hep-ex/9710027.

\bibitem{Chekanov:2002xk}
ZEUS Collaboration, S.~Chekanov {\em et~al.},
\newblock Eur.Phys.J. {\bf C27}, 531 (2003), hep-ex/0211040.

\bibitem{Chekanov:2006hv}
ZEUS Collaboration, S.~Chekanov {\em et~al.},
\newblock Nucl.Phys. {\bf B767}, 1 (2007), hep-ex/0604032.

\bibitem{Li:2020bub}
H.~T. Li, I.~Vitev, and Y.~J. Zhu,
\newblock JHEP {\bf 11}, 051 (2020), 2006.02437.

\bibitem{Ali:2020ksn}
A.~Ali, G.~Li, W.~Wang, and Z.-P. Xing,
\newblock Eur. Phys. J. C {\bf 80}, 1096 (2020), 2008.00271.

\bibitem{Li:2021txc}
H.~T. Li, Y.~Makris, and I.~Vitev,
\newblock Phys. Rev. D {\bf 103}, 094005 (2021), 2102.05669.

\bibitem{Liu:2022wop}
X.~Liu and H.~X. Zhu,
\newblock Phys. Rev. Lett. {\bf 130}, 091901 (2023), 2209.02080.

\bibitem{Liu:2023aqb}
H.-Y. Liu, X.~Liu, J.-C. Pan, F.~Yuan, and H.~X. Zhu,
\newblock Phys. Rev. Lett. {\bf 130}, 181901 (2023), 2301.01788.

\bibitem{Cao:2023rga}
H.~Cao, X.~Liu, and H.~X. Zhu,
\newblock (2023), 2303.01530.

\bibitem{Devereaux:2023vjz}
K.~Devereaux, W.~Fan, W.~Ke, K.~Lee, and I.~Moult,
\newblock (2023), 2303.08143.

\bibitem{Andres:2023xwr}
C.~Andres, F.~Dominguez, J.~Holguin, C.~Marquet, and I.~Moult,
\newblock (2023), 2303.03413.

\bibitem{Kang:2023oqj}
Z.-B. Kang, J.~Penttala, F.~Zhao, and Y.~Zhou,
\newblock (2023), 2311.17142.

\bibitem{Cao:2023qat}
H.~Cao, H.~T. Li, and Z.~Mi,
\newblock (2023), 2312.07655.

\bibitem{Stewart:2009yx}
I.~W. Stewart, F.~J. Tackmann, and W.~J. Waalewijn,
\newblock Phys.Rev. {\bf D81}, 094035 (2010), 0910.0467.

\bibitem{Stewart:2010tn}
I.~W. Stewart, F.~J. Tackmann, and W.~J. Waalewijn,
\newblock Phys.Rev.Lett. {\bf 105}, 092002 (2010), 1004.2489.

\bibitem{Kang:2012zr}
Z.-B. Kang, S.~Mantry, and J.-W. Qiu,
\newblock Phys.Rev. {\bf D86}, 114011 (2012), 1204.5469.

\bibitem{Kang:2013wca}
Z.-B. Kang, X.~Liu, S.~Mantry, and J.-W. Qiu,
\newblock Phys. Rev. D {\bf 88}, 074020 (2013), 1303.3063.

\bibitem{Kang:2013nha}
D.~Kang, C.~Lee, and I.~W. Stewart,
\newblock Phys. Rev. D {\bf 88}, 054004 (2013), 1303.6952.

\bibitem{Kang:2013lga}
Z.-B. Kang, X.~Liu, and S.~Mantry,
\newblock Phys. Rev. D {\bf 90}, 014041 (2014), 1312.0301.

\bibitem{Chu:2022jgs}
Z.~Chu, Y.~Wang, J.-H. Ee, J.~Chen, and D.~Kang,
\newblock JHEP {\bf 06}, 111 (2022), 2202.08040.

\bibitem{Kang:2014qba}
D.~Kang, C.~Lee, and I.~W. Stewart,
\newblock JHEP {\bf 11}, 132 (2014), 1407.6706.

\bibitem{Kang:2015swk}
D.~Kang, C.~Lee, and I.~W. Stewart,
\newblock PoS {\bf DIS2015}, 142 (2015).

\bibitem{Hessler:2021usr}
H1, J.~Hessler, D.~Britzger, and S.~Lee,
\newblock PoS {\bf EPS-HEP2021}, 367 (2022), 2111.11364.

\bibitem{Gehrmann-DeRidder:2016cdi}
A.~Gehrmann-De~Ridder, T.~Gehrmann, E.~W.~N. Glover, A.~Huss, and T.~A. Morgan,
\newblock JHEP {\bf 07}, 133 (2016), 1605.04295.

\bibitem{Currie:2016ytq}
J.~Currie, T.~Gehrmann, and J.~Niehues,
\newblock Phys. Rev. Lett. {\bf 117}, 042001 (2016), 1606.03991.

\bibitem{Currie:2017tpe}
J.~Currie, T.~Gehrmann, A.~Huss, and J.~Niehues,
\newblock JHEP {\bf 07}, 018 (2017), 1703.05977,
\newblock [Erratum: JHEP 12, 042 (2020)].

\bibitem{Gehrmann:2019hwf}
T.~Gehrmann, A.~Huss, J.~Mo, and J.~Niehues,
\newblock Eur. Phys. J. C {\bf 79}, 1022 (2019), 1909.02760.

\bibitem{Knobbe:2023ehi}
M.~Knobbe, D.~Reichelt, and S.~Schumann,
\newblock JHEP {\bf 09}, 194 (2023).

\bibitem{Bruser:2018rad}
R.~Br\"user, Z.~L. Liu, and M.~Stahlhofen,
\newblock Phys. Rev. Lett. {\bf 121}, 072003 (2018), 1804.09722.

\bibitem{Banerjee:2018ozf}
P.~Banerjee, P.~K. Dhani, and V.~Ravindran,
\newblock Phys. Rev. D {\bf 98}, 094016 (2018), 1805.02637.

\bibitem{Ebert:2020unb}
M.~A. Ebert, B.~Mistlberger, and G.~Vita,
\newblock JHEP {\bf 09}, 143 (2020), 2006.03056.

\bibitem{Baranowski:2022vcn}
D.~Baranowski, A.~Behring, K.~Melnikov, L.~Tancredi, and C.~Wever,
\newblock JHEP {\bf 02}, 073 (2023), 2211.05722.

\bibitem{Baranowski:2022khd}
D.~Baranowski, M.~Delto, K.~Melnikov, and C.-Y. Wang,
\newblock Phys. Rev. D {\bf 106}, 014004 (2022), 2204.09459.

\bibitem{Chen:2020dpk}
W.~Chen, F.~Feng, Y.~Jia, and X.~Liu,
\newblock JHEP {\bf 22}, 094 (2020), 2206.12323.

\bibitem{Moult:2016fqy}
I.~Moult, L.~Rothen, I.~W. Stewart, F.~J. Tackmann, and H.~X. Zhu,
\newblock Phys. Rev. D {\bf 95}, 074023 (2017), 1612.00450.

\bibitem{Boughezal:2016zws}
R.~Boughezal, X.~Liu, and F.~Petriello,
\newblock JHEP {\bf 03}, 160 (2017), 1612.02911.

\bibitem{Ebert:2018lzn}
M.~A. Ebert {\em et~al.},
\newblock JHEP {\bf 12}, 084 (2018), 1807.10764.

\bibitem{Boughezal:2018mvf}
R.~Boughezal, A.~Isgr\`o, and F.~Petriello,
\newblock Phys. Rev. D {\bf 97}, 076006 (2018), 1802.00456.

\bibitem{Nagy:2005gn}
Z.~Nagy and Z.~Trocsanyi,
\newblock Phys. Lett. B {\bf 634}, 498 (2006), hep-ph/0511328.

\bibitem{Abelof:2016pby}
G.~Abelof, R.~Boughezal, X.~Liu, and F.~Petriello,
\newblock Phys. Lett. B {\bf 763}, 52 (2016), 1607.04921.

\bibitem{Bauer:2000ew}
C.~W. Bauer, S.~Fleming, and M.~E. Luke,
\newblock Phys.Rev. {\bf D63}, 014006 (2000), hep-ph/0005275.

\bibitem{Bauer:2000yr}
C.~W. Bauer, S.~Fleming, D.~Pirjol, and I.~W. Stewart,
\newblock Phys.Rev. {\bf D63}, 114020 (2001), hep-ph/0011336.

\bibitem{Bauer:2001ct}
C.~W. Bauer and I.~W. Stewart,
\newblock Phys.Lett. {\bf B516}, 134 (2001), hep-ph/0107001.

\bibitem{Bauer:2001yt}
C.~W. Bauer, D.~Pirjol, and I.~W. Stewart,
\newblock Phys.Rev. {\bf D65}, 054022 (2002), hep-ph/0109045.

\bibitem{Bauer:2002nz}
C.~W. Bauer, S.~Fleming, D.~Pirjol, I.~Z. Rothstein, and I.~W. Stewart,
\newblock Phys.Rev. {\bf D66}, 014017 (2002), hep-ph/0202088.

\bibitem{Beneke:2002ph}
M.~Beneke, A.~Chapovsky, M.~Diehl, and T.~Feldmann,
\newblock Nucl.Phys. {\bf B643}, 431 (2002), hep-ph/0206152.

\bibitem{Manohar:2003vb}
A.~V. Manohar,
\newblock Phys.Rev. {\bf D68}, 114019 (2003), hep-ph/0309176.

\bibitem{Idilbi_2006}
A.~Idilbi, X.~Ji, and F.~Yuan,
\newblock Nuclear Physics B {\bf 753}, 42 (2006).

\bibitem{Becher:2006mr}
T.~Becher, M.~Neubert, and B.~D. Pecjak,
\newblock JHEP {\bf 01}, 076 (2007), hep-ph/0607228.

\bibitem{Bosch:2004th}
S.~Bosch, B.~Lange, M.~Neubert, and G.~Paz,
\newblock Nucl.Phys. {\bf B699}, 335 (2004), hep-ph/0402094.

\bibitem{Becher:2006qw}
T.~Becher and M.~Neubert,
\newblock Phys. Lett. B {\bf 637}, 251 (2006).

\bibitem{Becher:2010pd}
T.~Becher and G.~Bell,
\newblock Phys. Lett. B {\bf 695}, 252 (2011), 1008.1936.

\bibitem{Stewart:2010qs}
I.~W. Stewart, F.~J. Tackmann, and W.~J. Waalewijn,
\newblock JHEP {\bf 1009}, 005 (2010), 1002.2213.

\bibitem{Mantry:2009qz}
S.~Mantry and F.~Petriello,
\newblock Phys.Rev. {\bf D81}, 093007 (2010), 0911.4135.

\bibitem{Berger:2010xi}
C.~F. Berger, C.~Marcantonini, I.~W. Stewart, F.~J. Tackmann, and W.~J.
  Waalewijn,
\newblock JHEP {\bf 1104}, 092 (2011), 1012.4480.

\bibitem{Gaunt:2014xga}
J.~Gaunt, M.~Stahlhofen, and F.~Tackmann,
\newblock JHEP {\bf 04}, 113 (2014).

\bibitem{Gaunt:2014cfa}
J.~Gaunt, M.~Stahlhofen, and F.~J. Tackmann,
\newblock JHEP {\bf 08}, 020 (2014), 1405.1044.

\bibitem{Jouttenus:2011wh}
T.~T. Jouttenus, I.~W. Stewart, F.~J. Tackmann, and W.~J. Waalewijn,
\newblock Phys.Rev. {\bf D83}, 114030 (2011), 1102.4344.

\bibitem{Boughezal:2015eha}
R.~Boughezal, X.~Liu, and F.~Petriello,
\newblock Phys. Rev. D {\bf 91}, 094035 (2015), 1504.02540.

\bibitem{Henn:2019swt}
J.~M. Henn, G.~P. Korchemsky, and B.~Mistlberger,
\newblock JHEP {\bf 04}, 018 (2020), 1911.10174.

\bibitem{Moult:2022xzt}
I.~Moult, H.~X. Zhu, and Y.~J. Zhu,
\newblock (2022), 2205.02249.

\bibitem{Cacciari:2011ma}
M.~Cacciari, G.~P. Salam, and G.~Soyez,
\newblock Eur. Phys. J. C {\bf 72}, 1896 (2012), 1111.6097.

\bibitem{Kripfganz:1991}
J.~Kripfganz, H.~J. M{\"o}hring, and H.~Spiesberger,
\newblock Zeitschrift f{\"u}r Physik C Particles and Fields {\bf 49}, 501
  (1991).

\bibitem{BLUMLEIN2003242}
J.~Bl{\"u}mlein and H.~Kawamura,
\newblock Physics Letters B {\bf 553}, 242 (2003).

\bibitem{Afanasev:2004}
A.~V. Afanasev, I.~Akushevich, and N.~P. Merenkov,
\newblock Journal of Experimental and Theoretical Physics {\bf 98}, 403 (2004).

\bibitem{Liu:2020rvc}
T.~Liu, W.~Melnitchouk, J.-W. Qiu, and N.~Sato,
\newblock Phys. Rev. D {\bf 104}, 094033 (2021), 2008.02895.

\bibitem{Bauer:2003di}
C.~W. Bauer, C.~Lee, A.~V. Manohar, and M.~B. Wise,
\newblock Phys.Rev. {\bf D70}, 034014 (2004), hep-ph/0309278.

\bibitem{Ligeti:2008ac}
Z.~Ligeti, I.~W. Stewart, and F.~J. Tackmann,
\newblock Phys.Rev. {\bf D78}, 114014 (2008), 0807.1926.

\bibitem{Abbate:2010xh}
R.~Abbate, M.~Fickinger, A.~H. Hoang, V.~Mateu, and I.~W. Stewart,
\newblock Phys. Rev. D {\bf 83} (2011).

\bibitem{Stewart:2011cf}
I.~W. Stewart and F.~J. Tackmann,
\newblock Phys.Rev. {\bf D85}, 034011 (2012), 1107.2117.

\bibitem{Bierlich:2022pfr}
C.~Bierlich {\em et~al.},
\newblock SciPost Phys. Codeb. {\bf 2022}, 8 (2022).

\bibitem{Hoang:2007vb}
A.~H. Hoang and I.~W. Stewart,
\newblock Phys.Lett. {\bf B660}, 483 (2008), 0709.3519.

\bibitem{Becher:2009cu}
T.~Becher and M.~Neubert,
\newblock Phys. Rev. Lett. {\bf 102}, 162001 (2009).

\bibitem{Becher:2009qa}
T.~Becher and M.~Neubert,
\newblock JHEP {\bf 06}, 081 (2009).

\bibitem{Korchemsky:1987wg}
G.~Korchemsky and A.~Radyushkin,
\newblock Nucl.Phys. {\bf B283}, 342 (1987).

\bibitem{Moch:2004pa}
S.~Moch, J.~Vermaseren, and A.~Vogt,
\newblock Nucl.Phys. {\bf B688}, 101 (2004), hep-ph/0403192.

\bibitem{Moch:2005id}
S.~Moch, J.~Vermaseren, and A.~Vogt,
\newblock JHEP {\bf 0508}, 049 (2005), hep-ph/0507039.

\bibitem{Tarasov:1980au}
O.~Tarasov, A.~Vladimirov, and A.~Y. Zharkov,
\newblock Phys.Lett. {\bf B93}, 429 (1980).

\bibitem{Larin:1993tp}
S.~Larin and J.~Vermaseren,
\newblock Phys.Lett. {\bf B303}, 334 (1993), hep-ph/9302208.

\bibitem{vanRitbergen:1997va}
T.~van Ritbergen, J.~A.~M. Vermaseren, and S.~A. Larin,
\newblock Phys. Lett. B {\bf 400}, 379 (1997).

\bibitem{Lee:2006nr}
C.~Lee and G.~F. Sterman,
\newblock Phys. Rev. D {\bf 75} (2007).

\bibitem{Mateu:2012nk}
V.~Mateu, I.~W. Stewart, and J.~Thaler,
\newblock Phys. Rev. D {\bf 87}, 014025 (2013).

\end{thebibliography}

\end{document}